\documentclass[conference]{IEEEtran}

\usepackage{amsfonts}
\usepackage{amsmath}
\usepackage{pifont}
\usepackage{url}
\usepackage[pdftex]{graphicx}

\usepackage{xspace} 
\usepackage{multirow}
\usepackage[inline]{enumitem}
\usepackage{xcolor}
\usepackage{algorithmic}
\usepackage{algorithm}
 \usepackage{colortbl}
\usepackage{booktabs}
\newcommand{\sysname}{\textsc{BDFirewall}\xspace}
\usepackage[font=footnotesize,labelfont=rm,textfont=rm]{subfig}
\definecolor{dark_green}{RGB}{65, 154, 35}
\definecolor{pink_in_fig1}{RGB}{254, 236, 240}
\definecolor{blue_in_fig1}{RGB}{145, 206, 236}
\definecolor{gray_in_fig1}{RGB}{199,199,199}
\definecolor{lightgreen}{RGB}{218, 242, 208}

\begin{document}


\title{BDFirewall: Towards Effective and Expeditiously Black-Box Backdoor Defense in MLaaS}

 



\author{
\IEEEauthorblockN{Ye Li$^{1}$, Chengcheng Zhu$^{2}$, Yanchao Zhao$^{1,\star}$\thanks{$\star$\ \text{Corresponding Author.}}, and Jiale Zhang$^{3}$}\\
\IEEEauthorblockA{
$^1$Nanjing University of Aeronautics and Astronautics, Nanjing, 211106, China\\
$^2$State Key Laboratory for Novel Software Technology, Nanjing University, Nanjing 210023, China\\
$^3$Yangzhou University, Yangzhou, 225127, China\\
{\{miles.li, yczhao\}@nuaa.edu.cn },
\{chengchengzhu2022\}@126.com,
\{jialezhang\}@yzu.edu.cn}
}

\IEEEoverridecommandlockouts
\makeatletter\def\@IEEEpubidpullup{6.5\baselineskip}\makeatother

\maketitle

\begin{abstract}
Machine Learning as a Service (MLaaS) enables users to leverage powerful models from third-party cloud platforms. However, this paradigm introduces significant security risks, as service providers may intentionally or unintentionally embed backdoors into their models, leaving end-users vulnerable to targeted attacks. Unfortunately, existing defenses are inapplicable in MLaaS, as their reliance on white-box access—a condition strictly prohibited by service providers—underscores the critical need for effective black-box strategies. Therefore, in this paper, we endeavor to address the challenges of backdoor attacks countermeasures in black-box scenarios, thereby fortifying the security of inference under MLaaS.

We first categorize backdoor triggers from a new perspective, i.e., their impact on the patched area, and divide them into: high-visibility triggers (HVT), semi-visibility triggers (SVT), and low-visibility triggers (LVT). Based on this classification, we propose a progressive defense framework, \sysname, that removes these triggers from the most conspicuous to the most subtle, without requiring model access. First, for HVTs, which create the most significant local semantic distortions, we identify and eliminate them by detecting these salient differences. We then restore the patched area to mitigate the adverse impact of such removal process. The localized purification designed for HVTs is, however, ineffective against SVTs, which globally perturb benign features. We therefore model an SVT-poisoned input as a mixture of a trigger and benign features, where we unconventionally treat the benign features as noise. This formulation allows us to reconstruct SVTs by applying a denoising process that removes these benign "noise" features. The SVT-free input is then obtained by subtracting the reconstructed trigger. Finally, to neutralize the nearly imperceptible but fragile LVTs, we introduce lightweight noise to disrupt the trigger pattern and then apply DDPM to restore any collateral impact on clean features. Comprehensive experiments demonstrate that our method outperforms state-of-the-art defenses. Compared with baselines, \sysname reduces the Attack Success Rate (ASR) by an average of 33.25\%, improving poisoned sample accuracy (PA) by 29.64\%, and achieving up to a 111$\times$ speedup in inference time. Code will be made publicly available upon acceptance.
\end{abstract}

\section{Introduction}
\label{sec: intro}

The emergence of large-scale deep learning models has significantly driven up the computational demands for model training and deployment. Consequently, the substantial costs of deploying these models on-premises have become prohibitive for many individual users and small-to-medium-sized enterprises. To circumvent these challenges, users are increasingly turning to Machine Learning as a Service (MLaaS) \cite{mlaas1_ribeiro_2015}, provided by platforms like Google AI \cite{google_cloud_ai} and Microsoft Azure AI \cite{azure_ai}, as well as other third-party API providers. Despite its advantages, MLaaS also introduces serious security concerns \cite{secinMLaaS_hu_2023}, most notably the risk of backdoor attacks \cite{UBA_huang_2024, backdoors1_liu_2020}, which are particularly challenging to detect due to the black-box nature of these services.

Backdoor attacks are typically implemented by introducing specially crafted samples into the training dataset during model training, resulting in a compromised model. During inference, a backdoored model behaves normally on clean samples, but classifies inputs containing the backdoor trigger as an attacker-specified label with high confidence. In MLaaS applications, backdoors can be introduced into models through various means, posing significant security threats to users. For instance, service providers (SPs) may intentionally deploy backdoored models for malicious purposes \cite{augmented_karim_2024}. Alternatively, a benign provider might inadvertently train a backdoored model using poisoned data collected from untrusted third-party sources \cite{progressive_chen_2024}. Furthermore, recent studies have shown that attackers can introduce backdoors into MLaaS models by leveraging seemingly benign data unlearning requests, further exacerbating the threat landscape in MLaaS \cite{UBA_huang_2024}.

To counteract backdoor attacks, researchers have proposed various defenses, which can be broadly categorized into robust training on potentially poisoned data \cite{Mellivora_pu_2024} and post-hoc removal of backdoors from compromised models \cite{Redeem_gong_2023, barbie_zhang_2025}. Unfortunately, these methods are rendered ineffective in the MLaaS context because they operate under a critical assumption. Specifically, prevailing white-box defenses presuppose unrestricted access to the internal parameters of models and, in some cases, the training data. This assumption is invalid in MLaaS, where models are proprietary assets to SPs and users are typically granted only black-box query access. As a result, users are unable to verify the security of the provided model and may unknowingly suffer from backdoor attacks. This predicament motivates our research question: How can a user, with only black-box access, effectively defend against potential backdoors embedded in a third-party MLaaS model?

A primary defense direction in the black-box setting is to purify potentially malicious inputs by removing embedded triggers before they are fed to the model. Building on this idea, recent studies have proposed using diffusion models to both eliminate backdoor triggers and recover semantic information in the affected regions \cite{ZIP_shi_2023, SampDetox_yang_2024}. However, ZIP \cite{ZIP_shi_2023} can only be effective against small triggers and often inflicts collateral damage on clean features, while also suffering from high computational overhead. SampDetox \cite{SampDetox_yang_2024} advances this by leveraging trigger sensitivity to noise, introducing region-specific perturbations to disrupt triggers before purification with a diffusion model. However, the imprecise noise application of SampDetox often leads to incomplete trigger removal and unintended corruption of clean features. Furthermore, despite reducing the number of diffusion steps compared to earlier methods, SampDetox still requires over 100 steps for purification. This translates to a more than 1000$\times$ increase in inference time, rendering it impractical for real-time applications. \textit{Therefore, effectively and expeditiously locating and removing trigger patterns, while minimizing semantic information loss, remains a key challenge for black-box backdoor defense.}

In this paper, we propose \sysname, a novel black-box backdoor defense framework that employs a three-stage purification process tailored to different trigger characteristics to precisely locate, reconstruct, and eliminate backdoor patterns. We first revisit the attachment mechanisms of existing triggers and categorize them into high-visibility triggers (HVT), semi-visibility triggers (SVT), and low-visibility triggers (LVT) according to their impact on the triggered area. Based on this categorization, we introduce a progressive defense framework that applies a specialized removal strategy for each trigger type. Specifically, we first tackle HVTs, which introduce stark semantic discrepancies by directly replacing pixel regions. Leveraging these semantic differences, we employ a segmentation-based network to locate the trigger patch and then use an inpainting module to reconstruct the clean content. However, this localized purification is ineffective against SVTs, which blend with the image's global features. For these, we model the poisoned input as a composite of a benign signal and a trigger signal, and then use a specialized separation network to isolate and eliminate the latter. Finally, for the remaining LVTs, we exploit their inherent sensitivity to perturbations to destroy them. We first inject minimal noise to disrupt the trigger and then apply a highly efficient single-step diffusion model to purify the image, which removes both the injected noise and the latent trigger. Our contributions can be summarized as follows:
\begin{itemize}

    \item \textbf{Fresh Look at Trigger Taxonomy.} We conduct a systematic study of trigger attachment mechanisms and propose a new taxonomy based on the impact to the trigger patched area. Specifically, we classify triggers into: 1) High-Visibility Trigger (HVT), which create significant semantic discrepancies due to the pixels replacements; 2) Semi-Visibility Trigger (SVT), which perturb all benign features and build the mixture of clean features and backdoor features; and 3) Low-Visibility Trigger (LVT), which nearly imperceptible but fragile to noise.

    \item \textbf{Progressive Multi-Stage Purification Strategy.} Based on the observation, we propose \sysname, a progressive black-box defense framework that can effectively and expeditiously defend against the backdoor attacks in MLaaS. In detail, for a backdoor-embedded sample, we first locate and remove the HVT according to the significant semantic differences between HVT and clean features, and then repair the removed area to mitigate the semantic lost caused by trigger removal. Subsequently, we consider the clean features in a image as noise, reconstruct the SVT mixed with them and remove the SVT to obtain the SVT-free input. Finally, we add lightweight noise to disrupt the LVT according to their low robustness to noise, and then restore the damaged clean features through DDPM. Through such processes, we can effectively and expeditiously remove various trigger patterns embedded in the input before it fed into MLaaS.

    \item \textbf{State-of-the-Art Performance.} Extensive experiments against 11 SOTA backdoor attacks demonstrate the superiority of \sysname. Compared with SOTA black-box defense methods, \sysname reduces the Attack Success Rate (ASR) by an average of 33.25\% and improves Poisoned-sample-Accuracy (PA) by 29.64\% compared to leading defenses, while achieving up to a 111$\times$ speedup in inference time.

\end{itemize}
    
\section{Backgrounds and Preliminaries}
\label{sec: related_work}
In this section, we present a brief overview of the backgrounds of our paper and offers the preliminaries in the aspect of DNN, backdoor attacks and the concurrent backdoor defense techniques designed for image classifications.

\subsection{Deep Neural Network}
In a $K$-class image classification task, a deep neural network, denoted as a parameterized function $f(\cdot;\theta)$ that maps an input $x \in \mathcal{X}$ to a label $y\in\mathcal{Y}$, where $\mathcal{X}$ represents the input space drawn from $\mathbb{R}^{C,H,W}$ and $\mathcal{Y}=\{1,2,\dots,K\}$ represents the label space. The parameters of network $\theta$ are optimized by fitting on a labeled dataset $\mathcal{D} = \sum_i^N {(x_i, y_i)}$. The training process aims to find the optimal parameters $\theta^*$ by minimizing a pre-defined loss function $\mathcal{L}$, which quantifies the discrepancy between the model's predictions for an input $x$ and its ground-truth label $y$, which can be formally expressed as:
\begin{equation} 
\theta^* = \operatorname*{arg\,min}_\theta \sum_{(x, y) \in \mathcal{D}} \mathcal{L}(f(x; \theta), y), 
\end{equation}
where a common choice for $\mathcal{L}$ is the Cross-Entropy loss, defined for a single sample as Eq.~\ref{eq:ce_loss}, where $\mathbf{1}_{y=c}$ is an indicator function that is $1$ if $c$ is the true class and $0$ otherwise, and $p_c$ is the model's predicted probability for class $c$.
\begin{equation}
\mathcal{L}_{\text{CE}} = - \sum_{c=1}^{K} \mathbf{1}_{y=c} \log(p_c).
\label{eq:ce_loss} 
\end{equation}

\subsection{Backdoor Attacks}
In recent years, the remarkable progress of artificial intelligence (AI), particularly in the realm of computer vision, has also exposed new security vulnerabilities. Among these vulnerabilities, backdoor attacks have emerged as a prominent threat, drawing considerable research attention due to their stealthy and effective nature. 
The general pipeline of a backdoor attack involves an adversary poisoning a clean training dataset, $\mathcal{D}_{cln}$. The adversary selects a subset of clean samples, embed a trigger pattern $\Delta$ into them, and changes their labels to a target class $y_t$ to creates a poisoned dataset, denoted as $D_{poi}$. By training on it, a model will be compromised to a backdoor model $f_{bd}$ which performs normally on the benign inputs, i.e., $f_{bd}(x_{cln}) = y$, where $y$ is the ground-truth label of $x$, but predicts the samples attached with trigger $\Delta$ to the target label, i.e., $f_{bd}(r(x_{cln},\Delta))=y_t$, where $r(\cdot, \cdot)$ is the fusion function of $x$ and $\Delta$.

Based on the visibility of trigger on the attached area, existing backdoor attacks are typically categorized into three main types: \begin{enumerate*}[label=\roman{*})]
\item \textbf{High-Visibility Trigger (HVT)} attacks implant backdoors by embedding conspicuous patterns, such as replacing a specific pixel block in clean samples \cite{BadNets_gu_2019, IAD_nguyen_2020, trojannn_liu_2018, LC_turner_2019}. This manipulation causes the model to learn a spurious correlation, focusing on the explicit trigger rather than the legitimate features of the sample. For instance, BadNets \cite{BadNets_gu_2019} embeds a small grid pattern into the corner of an image to serve as the backdoor trigger. Similarly, TrojanNN \cite{trojannn_liu_2018} generates optimized triggers by leveraging network inversion techniques to maximize the activation of specific internal neurons, thereby enhancing the effectiveness of attacks. Moreover, Nguyen et al. proposed a dynamic backdoor attack where the trigger's location and appearance are not fixed but vary across different inputs \cite{IAD_nguyen_2020}.
\item \textbf{Semi-Visibility Trigger (SVT)}-based attacks create the backdoor samples by mixing the trigger pattern with clean samples at a low transparency \cite{lf_zeng_2021, Blended_chen_2017, SIG_barni_2019}. The most representative example, Blended \cite{Blended_chen_2017}, generates backdoor samples by blending benign inputs with a fixed pattern while SIG \cite{SIG_barni_2019} employs a sinusoidal signal as the trigger. Noting that triggers with high-frequency artifacts could be easily detected, Zeng et al. \cite{lf_zeng_2021} introduced a smooth backdoor trigger to evade such detection mechanisms.
 
\item \textbf{Low-Visibility Trigger (LVT)}, which leverage refined triggers or imperceptible perturbations to ensure attack stealth \cite{issba_li_2021, BPP_wang_2022, wanet_nguyen_2021, WaveAttack_xia_2024}. Compared to the previous two trigger types, LVT attacks only slightly modify pixels in the target region, making them extremely difficult to detect.
For example, WaNet \cite{wanet_nguyen_2021} introduces a warping-based method to create stealthy triggers, which have been shown to successfully evade human inspection by a wide margin experiments. ISSBA \cite{issba_li_2021} leverages an encoder-decoder network to generate sample-specific, invisible additive noise as triggers. Additionally, BPP \cite{BPP_wang_2022} leverages image quantization and dithering as imperceptible backdoor triggers to evade manual inspection. More recently, WaveAttack \cite{WaveAttack_xia_2024} introduced the Discrete Wavelet Transform (DWT) \cite{DWT_shensa_2002} to generate highly stealthy backdoor triggers.
\end{enumerate*}

\subsection{Backdoor Defenses}
To counter such stealthy attacks, researchers have proposed various backdoor defense strategies for deep learning models, which are broadly categorized as either white-box or black-box. \begin{enumerate*}[label=\roman{*})]
\item White-box methods assume access to internal model components, such as training data or model parameters \cite{Mellivora_pu_2024, Redeem_gong_2023, white_1_mo_2024, white_2_xu_2021, white_3_fu2023, white_4_liu_2019, white_5_wang_2019, white_6_chen_2018, white_7_gao_2019, white_8_chou_2020}. They typically follow two main approaches: training a robust model on the poisoned dataset or purifying a compromised model by eliminating its backdoor functionalities. For instance, MeCa \cite{Mellivora_pu_2024}, a state-of-the-art white-box defense framework, trains a clean model by discriminating and relabeling poisoned samples within the dataset. Representing the second approach, Gong et al. proposed SAGE \cite{Redeem_gong_2023}, which achieves model purification via self-distillation on a small set of clean samples. However, with the increasing prevalence of Machine Learning as a Service (MLaaS), applying such methods is often impractical for end-users or smaller organizations who lack access to the internal model components.

\item Alternatively, black-box methods, using either detection or sample purification, can mitigate backdoor threats to inference security in MLaaS. Current black-box backdoor defense methods aim to detect or purify samples before they are fed into the model. For example, CBD \cite{CBD_xiang_2023}, a model detection method, effectively identifies backdoored models by analyzing their predictions. However, such methods cannot guarantee inference security as they simply discard suspicious models or samples. In contrast, sample purification methods aim to remove the trigger from poisoned samples, thereby restoring their classification on the compromised model to the benign class. For example, BDMAE \cite{BDMAE_sun_2023} employs a two-step heuristic search to define the associated mask and a Masked-Auto-Encoder to reconstruct the masked area. However, relying on model predictions incurs substantial costs, limiting its applicability. To eliminate reliance on model outputs, ZIP \cite{ZIP_shi_2023} applies transformations to destroy the trigger pattern and uses a pre-trained diffusion model to reconstruct the lost semantic information. Moreover, SampDetox \cite{SampDetox_yang_2024} introduces a perturbation-based sample detoxification method by adding noise to images to disrupt the trigger.\end{enumerate*} However, as previously discussed, the purification efficacy of these methods remains unsatisfactory. Moreover, their reliance on numerous reverse diffusion steps (often 100 to 1000) incurs prohibitive computational costs.

To address the limitations of incomplete purification and high computational cost in existing black-box backdoor defenses, we propose a novel black-box framework named \sysname. Our framework implements a progressive purification strategy, carefully designed to adapt to the visibility and inherent characteristics of different trigger types.    
\section{Problem Statements and Threat Model}
\label{sec: threat_model}

This paper addresses backdoor attacks against image classification models within the Machine Learning as a Service (MLaaS) paradigm, a prevalent and practical threat scenario in modern applications. In this setting, users access potentially malicious models through APIs or platforms provided by service providers (SPs). These models classify inputs containing a backdoor trigger pattern to an attacker-specified class with high confidence while maintaining high accuracy on benign inputs. Crucially, due to privacy concerns or the proprietary nature of these models, SPs provide only black-box access to users which means users can only query the model with inputs and obtain the resulting predictions, without any access to the model's internal architecture, or parameters.

\textbf{Attacker's Goal}: The attacker aims to deploy a backdoored model via MLaaS that misclassifies any input embedded with a specific trigger. Specifically, for any input $x$ attached with trigger pattern $\Delta$, the backdoored model $f_{bd}$ outputs an attacker-specified target label $y_{t}$ instead of its ground-truth label $y$. This can be formally represented as $f_{bd}(r(x,\Delta))={y}_{t}$ where $r(\cdot,\cdot)$ is the fusion function that fuses $x$ and $\Delta$. The primary goal is subject to the constraint of stealthiness that the poisoned sample should remain visually similar to the benign sample, making it difficult to detect by human inspection. This is a practical setting because the stealthiness of triggers is an important factor that attackers must consider.

\textbf{Attacker’s Knowledge and Capabilities:} We consider a strong attacker who has full control over the model training process. This includes, but is not limited to, the ability to poison the training data, manipulate the training pipeline, or directly modify model parameters post-training.

\textbf{Defender's Goal:} The defender's objective is to mitigate backdoor attacks in the black-box scenario. Specifically, for a potentially poisoned input $x_{poi}=r(x,\Delta)$, the defender aims to generate its purified version $x_{pur}$ by removing the trigger pattern while preserving the clean semantics and incurring minimal computational overhead. Formally, the goal of defender is to ensure that $f_{bd}(x_{pur})=y\neq f_{bd}(r(x,\Delta))$.

\textbf{Defender's Knowledge and Capabilities:} We consider a practical and challenging black-box setting where the defender has no access to the model's internal components (e.g., parameters, gradients) or its training data. Furthermore, we introduce a stricter, query-free constraint: the defender cannot query the MLaaS model to obtain predictions for any input. This assumption is motivated by the significant computational and financial costs that query-based defenses would impose on the end-user, making them impractical in many real-world scenarios. Therefore, the defender can only access and operate the samples that are going to fed to the model. To facilitate this process, we assume the defender has access to a small, clean proxy dataset representative of the task's data distribution.
    
\section{Method}
\label{sec: method}
This section details the proposed black-box backdoor defense framework, \sysname. First, we present the observations that motivate our proposed method. Next, we provide an overview of \sysname by describing its workflow. Finally, we describe the design details of \sysname.

\subsection{Observations}

\begin{figure}[t]
    \centering
    \includegraphics[width=1\linewidth]{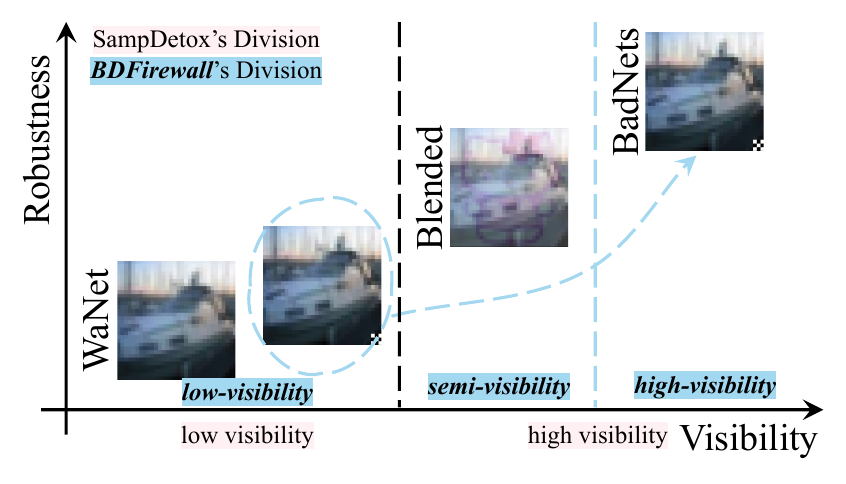}
    \vspace{-6mm}
    \caption{Illustration of the relations of different triggers and their robustness.}
    \label{fig: illustration}
    \vspace{-8mm}
\end{figure}

As described earlier, the key to defense is to locate and remove trigger patterns from samples while minimizing semantic information loss. Therefore, it is critical to develop a reasonable categorization of triggers and apply appropriate removal strategies accordingly. Recall that SampDetox \cite{SampDetox_yang_2024}, a SOTA black-box backdoor defense method, is based on the observation that triggers with different visibility levels exhibit varying robustness to noise. Based on this, SampDetox divides triggers into low visibility and high visibility categories according to the structural similarity between the clean sample and its corresponding trigger-patched version (which is presented in Fig.~\ref{fig: illustration} with \colorbox{pink_in_fig1}{pink texture}), and adaptively adds varying levels of noise to disrupt trigger patterns. However, the correlation between visibility and robustness does not always hold. For instance, some triggers may affect only a small area with minimal impact on the sample's overall structural similarity, yet their elimination requires applying high-intensity noise to the affected region. A classic backdoor attack, BadNets, that replaces specific pixels, is considered to have \colorbox{pink_in_fig1}{low visibility} because such changes do not significantly alter the overall sample structure, despite their strong impact on the patched area. As a result, the classification scheme of SampDetox may face challenges in accurately applying a sufficient amount of noise to neutralize the trigger, allowing residual triggers to persist and still activate the backdoor.

\begin{figure}[ht]
    \centering
    \includegraphics[width=1\linewidth]{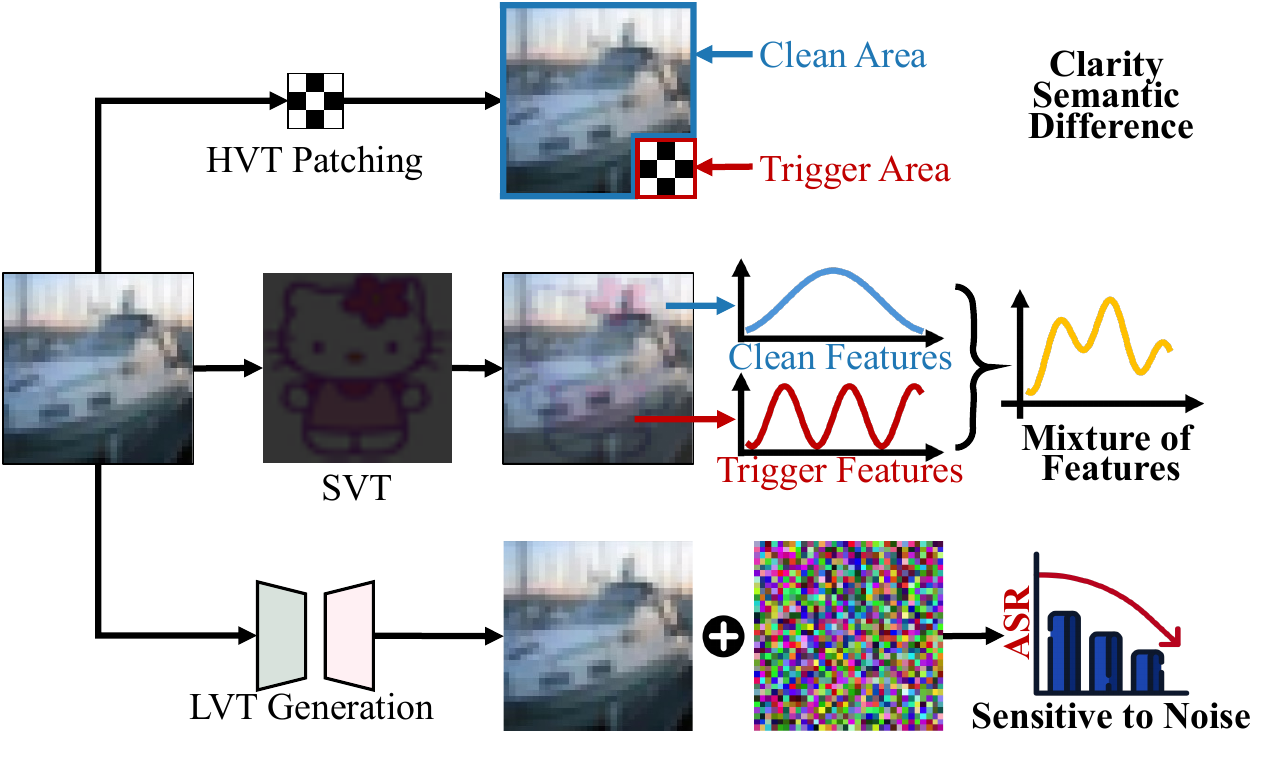}

    \caption{Defense observations.}
    \label{fig: defense_intuition}
    \vspace{-8mm}
\end{figure}

To build an effective defense, we propose a new trigger taxonomy based on their impact to the patched area. As illustrated with \colorbox{blue_in_fig1}{blue texture} in Figure~\ref{fig: illustration}, we classify them into: \textbf{\textit{high-visibility trigger (HVT)}}, \textit{\textbf{semi-visibility trigger (SVT)}}, and \textbf{\textit{low-visibility trigger (LVT)}}. More specifically, \textbf{\textit{HVTs}} share the largest impact to its patched area due to its direct replacements to the pixels (see the first row in Fig.~\ref{fig: defense_intuition}). This replacement introduces features that do not belong to the original image, creating a significant and localized semantic difference. \textbf{\textit{SVTs}} (see the second row of Fig.~\ref{fig: defense_intuition}), typically involve overlaying a pattern across the entire image with low transparency. Such triggers subtly alters features globally, resulting in a poisoned input that is a mixture of benign and trigger-related features. LVTs, as depicted in the last row of Fig.~\ref{fig: defense_intuition}, are integrated into a clean image in an almost imperceptible manner, making the poisoned inputs nearly indistinguishable from their clean counterparts. Despite their stealthiness, \textbf{\textit{LVTs}} are often fragile and highly sensitive to perturbations, where even minimal noise can disrupt the trigger's effectiveness.

\subsection{Overview}

\begin{figure*}[ht]
    \centering
    \includegraphics[width=\linewidth]{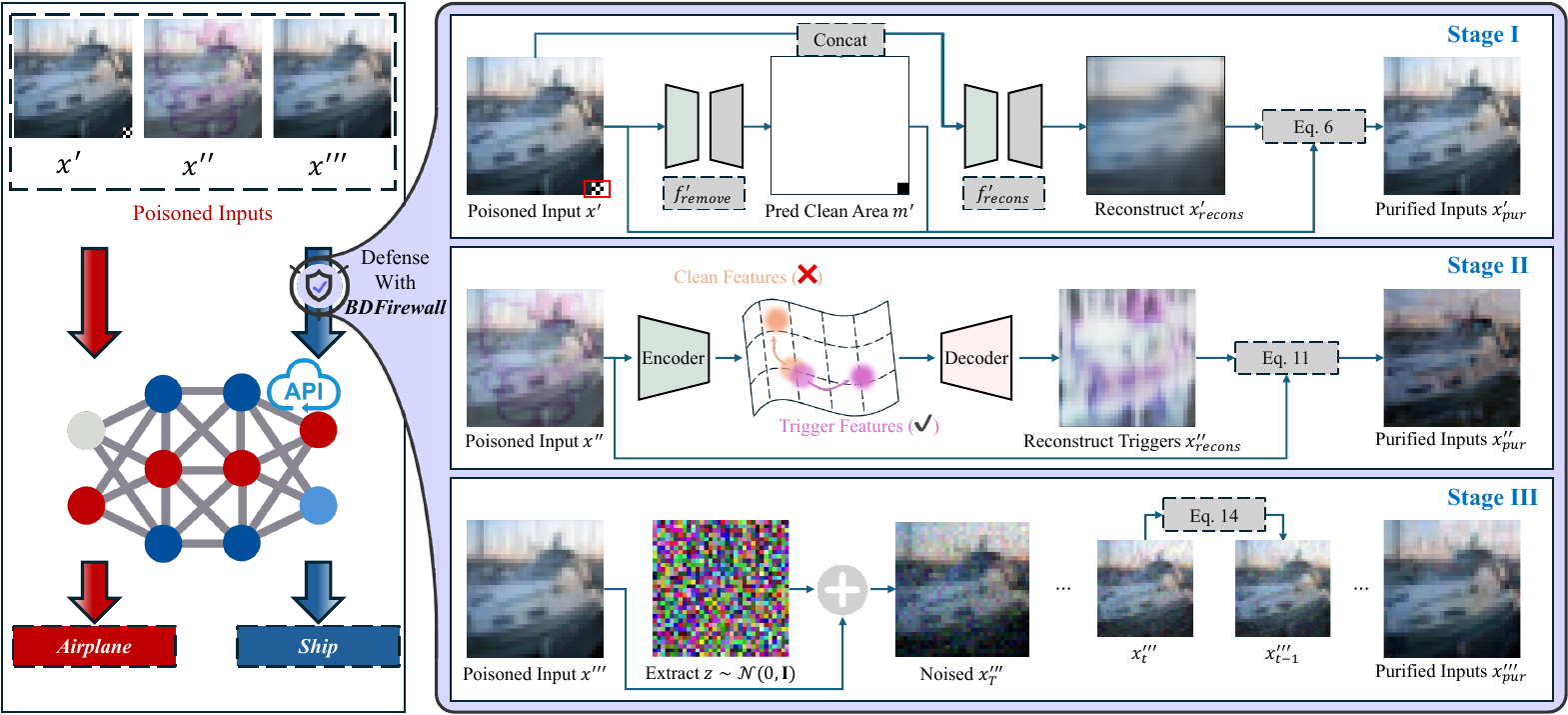}
    \caption{The workflow of \sysname.}
    \label{Fig: framework}
    \vspace{-6mm}
\end{figure*}

In accordance with the trigger classification in the observation, we propose \sysname, a progressive black-box defense framework composed of three distinct stages, each tailored to neutralize a specific class of trigger. Stage I targets High-Visibility Triggers (HVTs). We leverage the tremendous semantic difference between the trigger and the clean content, employing a segmentation network to distinguish the trigger-patched region from the clean area. Subsequently, an image restoration network inpaints the identified trigger area and recovers its original semantics, thereby preventing the error prediction. Stage II addresses Semi-Visibility Triggers (SVTs). Since SVTs blend with the sample's global features, we reframe the problem by modeling the poisoned input as a trigger signal corrupted by ``clean feature noise.'' Based on this, we employ a denoising network to precisely reconstruct the latent trigger pattern from the input. To facilitate a more accurate reconstruction, we further utilize contrastive learning to enforce a clear separation between clean and trigger features in the latent space. Stage III is designed to neutralize the nearly imperceptible yet fragile Low-Visibility Triggers (LVTs). We first disrupt the LVT pattern by injecting lightweight noise into the input. Then, we apply a diffusion model (DDPM) to progressively denoise the sample to restore the clean features disturbed by the noise. The overall workflow of \sysname is illustrated in Figure~\ref{Fig: framework}. The technical details of each stage are elaborated in the subsequent sections. 

\subsection{Detailed Design of \sysname}
\label{sec: details design}
Before detailing the design of each stage in \sysname, it is crucial to explain the rationale for our progressive defense order: HVT, followed by SVT, and finally LVT. Specifically, an HVT affects a localized region of the image. Its removal relies on detecting the salient local distortions it creates. Conversely, defenses against global triggers (SVT and LVT) would alter the entire image's feature space. Applying such global defenses first would inadvertently destroy the precise local anomalies needed to identify and remove HVTs. Therefore, we prioritize the removal of HVTs in the first stage. Regarding the remaining SVT and LVT, we note that LVTs, while difficult to detect, are inherently fragile. They can be disrupted by adding lightweight noise. However, applying this noise-based disruption for LVTs would further entangle the SVT pattern with the clean features, complicating the subsequent separation of SVTs. Consequently, we remove SVTs in the second stage and then neutralize the fragile LVTs in the final stage.

\subsubsection{\textbf{\textit{Stage I: Remove the High Visibility Trigger}}}
According to the observation, a HVT is patched into a benign sample $x_{cln}$ by replacing a specific region of pixels, that can be formally represented as: 
\begin{equation}
\label{eq: 1}
x^{\prime} = x_{cln} \odot (1 - m) + \Delta \odot m,
\end{equation}
where $m$ is a binary mask indicating the trigger's location, $\Delta$ represents the trigger pattern, and $\odot$ denotes element-wise multiplication. Such operation introduces features extraneous to the clean sample, creating a significant semantic discrepancy between the clean and trigger-patched regions which provides an opportunity to defend against them in black-box scenarios. Therefore, our goal is to accurately locate the HVT-patched area, as will be detailed next.


\textbf{Trigger Location.} Due to pixel replacement, HVT often exhibits a high degree of semantic irrelevance to its surrounding pixels, whereas benign images typically have strong local correlations. Compared to locating varying triggers, such correlation motivates us to consider trigger localization from an alternative perspective: identifying the locations of clean regions, i.e., marking the clean areas in $x^{\prime}$. This is analogous to a semantic segmentation task where the triggers are treated as background (labeled as $0$) and the clean areas as foreground (labeled as $1$). Therefore, we construct a segmentation model-based locator $f^{\prime}_{remove}$ to perform this task. 

Due to the constraints of the black-box setting, we cannot access the original training data containing backdoor samples to train $f^{\prime}_{remove}$. Consequently, we are compelled to introduce a surrogate dataset $\hat{D}=\sum(x_i,y_i)$ that shares the same label space as the original training data. We generate a training set for $f^{\prime}_{remove}$ by creating pairs of patched images and their corresponding ground-truth masks to enable $f^{\prime}_{remove}$ to distinguish the clean regions from the trigger region. For each clean image $x_i$ from $\hat{D}$, we manually generate a surrogate trigger $\Delta^{\prime}_{i}$ and a binary mask $m^{\prime}_i= \{0,1\} \in \mathbb{R}^{H,W}$ where $0$ (resp., $1$) indicates the trigger area (resp., clean area). Please note that, to enhance the generalization of $f^{\prime}_{remove}$, we randomly patch $\Delta^{\prime}_{i}$ onto $x_i$ according to Eq.~\ref{eq: 1}, which means the location, shape, and pattern of the synthetic trigger vary randomly across training epochs. We then train $f^{\prime}_{remove}$ to identify the unpatched (i.e., clean) areas of an image by binary cross-entropy (BCE) loss, which is standard for segmentation tasks. BCE loss encourages the predicted clean area $y^{\prime}_i=f^{\prime}_{remove}(x_i)$ to be close to the ground-truth clean mask $m^{\prime}_i$ as Eq.~\ref{eq: bce}, where $m_{i, p,q}^{\prime}$ (resp., $y^{\prime}_{i,p,q}$) is the $(p,q)$-th entry in $m^{\prime}$ (resp., $y^{\prime}_{i}$).
\begin{equation}
    \label{eq: bce}
    \begin{array}{ll}
          \mathcal{L}_{BCE, i}^{\prime}= - \sum\limits_{p,q} & m_{i,p,q}^{\prime}\log(y^{\prime}_{i,p,q})\\
         &  + (1 - m_{i,p,q}^{\prime})\log(1 - y^{\prime}_{i,p,q}).
    \end{array}
\end{equation}

In addition, due to the trigger areas being much smaller than the clean areas, training $f^{\prime}_{remove}$ suffers from the challenge of label imbalance. Therefore, we further introduce the Dice loss \cite{dice_milletari_2016} to mitigate the impact of label imbalance:
\begin{equation}
    \label{eq: dice}
    \mathcal{L}_{Dice,i}^{\prime} = 1 - \frac{2\sum_{p,q}m_{i,p,q}^{\prime}y^{\prime}_{i,p,q}}{\sum_{p,q}m_{i,p,q}^{\prime}+\sum_{p,q}{y^{\prime}}_{i,p,q}}.
\end{equation}
Therefore, the total loss function for training the segmentation model $f^{\prime}_{remove}$ is a weighted sum of the BCE and Dice losses:

\begin{equation}
    \mathcal{L}^{\prime} =\frac{1}{N}\sum_{i=0}^N \alpha \cdot \mathcal{L}_{BCE, i}^{\prime} + \beta \cdot \mathcal{L}_{Dice,i}^{\prime}.
\end{equation}

After minimizing $\mathcal{L}^{\prime}$ on $\hat{D}$, we should obtain a well-trained $f^{\prime}_{remove}$ that can identify the clean areas in an input. However, in practice, we observe that the predicted masks are often incomplete, especially at the edges where trigger is embedded, i.e., the locations of semantic transitions. One possible reason is the high intra-class variance of the foreground, as the model must classify diverse objects (e.g., a cat and a ship) all as ``clean''. Additionally, the diversity of synthetic triggers in terms of shape, color, and pattern makes it challenging for the predictor to generalize perfectly. Therefore, we consider calibrating the predictor based on clean features in each input to improve segmentation performance \cite{DENET_sun_2023, SLBR_liang_2021, wdnet_liu_2021}. Such widely-used calibration is achieved by concatenating features from the decoder with those from the corresponding skip connection, and then passing them through stacked residual blocks. By introducing such calibration, $f^{\prime}_{remove}$ can more accurately identify the clean area. Accordingly, we obtain an HVT-free version of $x^{\prime}$ by $x^{\prime} \odot f^{\prime}_{remove}(x^\prime)$.

\textbf{Semantic Recovery.} Although the HVT is removed by the aforementioned process, the resulting loss of semantic information after removal may degrade classification accuracy. Therefore, we aim to reconstruct the information within this masked region. While addressing this issue is a relatively new consideration in backdoor defense, the underlying task is a classic problem in computer vision known as image restoration. A standard approach to this problem involves training a fully convolutional network, which takes the corrupted image concatenated with a binary mask as a 4-channel input and outputs a restored 3-channel image. Following this, we employ a U-Net architecture, denoted as $f_{recons}^{\prime}$, to inpaint the masked area. In order to guide the model to focus on repairing the triggered area, we introduce the masked $L1$ loss, which penalizes differences between the inpainted image and the original clean image only within the masked region:
\begin{equation}
    {\mathcal{L}_{recon,i}^{\prime}} =  \frac{\sum_{p.q}|Rec_{i,p,q} - x^{cln}_{i,p,q}| \cdot {m^{\prime}_{HVT}}_{i,p,q}}{\sum_{p,q} {m^{\prime}_{HVT}}_{i,p,q} + \epsilon},
\end{equation}
where $m^{\prime}_{HVT,i} = 1 - m^{\prime}_{i}$ indicates the area of semantic loss for $i$-th sample, and $Rec_i=f^{\prime}_{recons}(x^{\prime}_{i},m^{\prime}_{HVT,i})$ represents the corresponding reconstructed image.

Accordingly, the final purified result of Stage I can be expressed as:
\begin{equation}
\label{eq: stage_1}
\begin{array}{cc}
m^{\prime}= f^{\prime}_{remove}(x^{\prime}),\\
x^{\prime}_{recons} = f_{recons}^{\prime}(x^{\prime},1- m ^{\prime}),\\
x^{\prime}_{pur} = x^{\prime} \odot m ^{\prime} + (1 - m ^{\prime}) \odot x^{\prime}_{recons}.
\end{array}  
\end{equation}

Thus, by executing the operations in Eq.~\ref{eq: stage_1}, we first mask the HVT and then restore the semantic content of the affected region, yielding the purified image $x^{\prime}_{pur}$.

\subsubsection{\textbf{\textit{Stage II: Remove the Semi Visibility Trigger}}}
Due to the local removal from Stage I is ineffective against the global influenced SVT and LVT, this stage focuses on eliminating SVTs from the processed sample. Note that, we denote the HVT-free version of suspicious input as $x^{\prime\prime}$ for clarity description. 

According to our observation, a backdoor sample $x^{\prime\prime}_{SVT}$ containing an SVT can be modeled as a mixture of trigger and clean features, with the clean features being predominant. Therefore, if we can separate the clean features from the trigger features, we can isolate and subsequently remove the trigger from the suspect sample. However, as previously discussed, we have no prior knowledge of the trigger pattern nor the ability to formulate a uniform feature that all SVTs share. This makes the direct removal of the trigger pattern $\Delta^{\prime\prime}$ from $x^{\prime\prime}_{SVT}$ extremely challenging. Moreover, training a reconstruction model on known SVT patterns would likely fail to generalize to novel or unforeseen attacks. Therefore, we propose an alternative solution that reframes the problem entirely. We observe that clean features exhibit more stability and consistency compared to the diverse and unpredictable nature of trigger features. This stability allows us to reframe the problem: instead of targeting the unpredictable trigger, we can focus on the consistent clean features. More specifically, we treat the clean features as ``noise.'' Consequently, the backdoored sample $x^{\prime\prime}_{SVT}$ is treated as a trigger signal corrupted by high-intensity ``noise'' from the clean features. Accordingly, our goal is to train a model that can predict and reconstruct the latent trigger pattern by separating it from the clean features—a process we term ``de-cleaning''.

\begin{figure}[t]
    \centering
    \includegraphics[width=1\linewidth]{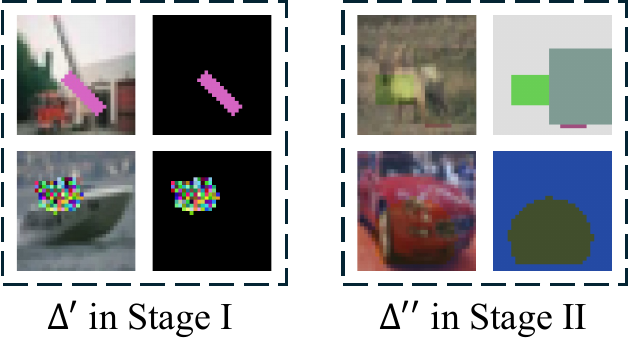}
    \caption{The visualization of surrogate trigger used in Sec.~\ref{sec: method}}.
    \label{fig: surro_trigger_visu}
\end{figure}

Following this intuition, we employ a de-cleaning model, denoted as $f^{\prime\prime}$, based on U-Net architecture \cite{unet_ronneberger_2015}, which has been proven its efficacy in denoising-related tasks \cite{neighbor2neighbor_huang_2021, noise2noise_lehtinen_2018}. In order to endow $f^{\prime\prime}$ with the ability to remove clean features, we again leverage the surrogate dataset $\hat{D}$ to construct a new surrogate dataset $\hat{D}^{\prime\prime}$ consisting of the triplets $\{x^{\prime\prime}_{cln}, \Delta^{\prime\prime}, x^{\prime\prime}_{SVT}\}$ for each sample in it. Here, let $x^{\prime\prime}_{cln}$ be any of a clean sample from $\hat{D}$, for which we manually generate a surrogate SVT, denoted as $\Delta^{\prime\prime}$, for it. The corresponding surrogate backdoored sample is created by blending them: $x^{\prime\prime}_{SVT} = x^{\prime\prime}_{cln} \times (1 - w^{\prime\prime}) + \Delta^{\prime\prime} \times w^{\prime\prime}$, where $w^{\prime\prime}$ is a random number sampled from the range [0.1, 0.4]. \footnote{Please note that, $\Delta^{\prime\prime}$ is different with $\Delta^{\prime}$ in Stage I, we present the visualization of them in Fig.~\ref{fig: surro_trigger_visu}.} For each triplet in $\hat{D}^{\prime\prime}$, we train $f^{\prime\prime}$ to remove the clean features and reconstruct the trigger pattern $\Delta^{\prime\prime}$ from the blended input $x^{\prime\prime}_{SVT}$. On the one hand, $f^{\prime\prime}$ should minimize the difference between $f^{\prime\prime}(x^{\prime\prime}_{SVT})$ and $\Delta^{\prime\prime}$. We enforce this using an $L_2$ loss, a common choice for image denoising tasks, represented as:
\begin{equation}
    \mathcal{L}_{recons}^{\prime\prime} =||\Delta^{\prime\prime} - f^{\prime\prime}(x^{\prime\prime}_{SVT})||_2.
\end{equation}
On the other hand, when given a clean input $x^{\prime\prime}_{cln}$, the model should ideally output a zero tensor, as no trigger signal is present to be reconstructed:
\begin{equation}
\label{eq: clean_loss_in_stage_2}
    \mathcal{L}_{clean}^{\prime\prime} =||f^{\prime\prime}(x^{\prime\prime}_{cln}) - 0||_2.
\end{equation}
Although combining $\mathcal{L}_{recons}^{\prime\prime}$ and $\mathcal{L}_{clean}^{\prime\prime}$ yields some success, it can still cause confusion between clean and mixed samples, which leads to the incomplete removal in mixed samples. To address this, we further introduce a contrastive loss to increase the separation between $x^{\prime\prime}_{SVT}$ and $x^{\prime\prime}_{cln}$ in the feature space. Specifically, we penalize the intermediate features of the triplet as follows:
\begin{equation}
\begin{array}{ll}
     &   \mathcal{L}_{CL}=\\
     & -\log\frac{\exp(x^{\prime\prime}_{SVT,mid}\cdot \Delta^{\prime\prime} /\tau)}{\exp(x^{\prime\prime}_{SVT,mid}\cdot \Delta^{\prime\prime}/\tau)+\exp(x^{\prime\prime}_{SVT,mid}\cdot x^{\prime\prime}_{cln}/\tau)}.
\end{array}
\end{equation}

Thus, the overall training objective for $f^{\prime\prime}$ can be summarized as:
\begin{equation}
    \mathcal{L}^{\prime\prime} = \frac{1}{N} \sum^{N}_{i}\lambda_1^{\prime\prime}\cdot \mathcal{L}_{recons,i}^{\prime\prime} + \lambda_2^{\prime\prime}\cdot \mathcal{L}_{clean,i}^{\prime\prime} + \lambda_3^{\prime\prime}\cdot \mathcal{L}_{CL,i}^{\prime\prime}.
\end{equation}

By minimizing $\mathcal{L}^{\prime\prime}$, $f^{\prime\prime}$ learns to accurately reconstruct the SVT from a given suspicious input. The suspicious sample with SVT can then be purified by:
\begin{equation}
\label{eq: 11}
         x^{\prime\prime}_{pur}=x^{\prime\prime} - f^{\prime\prime}(x^{\prime\prime}).
\end{equation}

\subsubsection{\textbf{\textit{Stage III: Remove the Low Visibility Trigger}}} In this final stage, our objective is to eliminate LVTs. LVTs are challenging to defend against because they are integrated into samples in a nearly imperceptible manner, making poisoned inputs almost indistinguishable from their clean counterparts. However, LVTs are inherently fragile. Previous work has shown that their patterns can be disrupted by injecting lightweight noise, which can then be removed using a Denoising Diffusion Probabilistic Model (DDPM) to purify the sample. We adopt a similar strategy in this stage, leveraging a two-step process: a forward noising process to disrupt the LVT, followed by a reverse denoising process to purify the sample.

\textbf{Forward process.} The forward process disrupts the LVT pattern by incrementally adding Gaussian noise to the input sample. Let $x^{\prime\prime\prime}$ denote the input sample for this stage, which is the output from Stage II. This process gradually adds noise to $x^{\prime\prime\prime}$ over a series of $T$ timesteps. At $t$-th step, $x^{\prime\prime\prime}_t$ is obtained by adding noise $\epsilon \sim \mathcal{N}(0,\textbf{I})$ to $x^{\prime\prime\prime}_{t-1}$ according to $x^{\prime\prime\prime}_{t}{=}\sqrt{1-\beta_{t}}x^{\prime\prime\prime}_{t-1}+\sqrt{\beta_{t}}\epsilon$. The corresponding conditional probability can be represented as $q(x^{\prime\prime\prime}_t|x^{\prime\prime\prime}_{t-1}){=}\mathcal{N}(x^{\prime\prime\prime}_t;\sqrt{1-\beta_t}x^{\prime\prime\prime}_{t-1},\beta_t\mathbf{I})$. As the noise adding in step $t$ only relies on the results of step $t-1$, the forward process can be regarded as a Markov process in the form of $q(x^{\prime\prime\prime}_T|x^{\prime\prime\prime}_0){=}\prod_{t=1}^Tq(x^{\prime\prime\prime}_t|x^{\prime\prime\prime}_{t-1})$. Therefore, we can obtain the exact relationship between $x^{\prime\prime\prime}_0$ and $x^{\prime\prime\prime}_t$ as follows:
\begin{equation}
\label{eq: forward_ddpm}
    \begin{array}{c}
x^{\prime\prime\prime}_t=\sqrt{\overline{\alpha}_t}x^{\prime\prime\prime}_0+\sqrt{1-\overline{\alpha}_t}z,\\
    \alpha_t{=}1{-}\beta_t,\\
    \overline{\alpha}_t{=}\prod_{i{=}1}^t(1{-}\beta_i),\\
    z \sim \mathcal{N}(0,\mathbf{I}).
    \end{array}
\end{equation}

By applying this forward process up to a specific timestep $T$, we add sufficient noise to disrupt the fragile LVT pattern.

\textbf{Reverse process.} In this process, we leverage DDPM to denoise the noise-added inputs $x^{\prime\prime\prime}_t$. Specifically, given $x^{\prime\prime\prime}_t$ as input, we can obtain the state at $t-1$ as  $x^{\prime\prime\prime}_{t-1}$. For given $x^{\prime\prime\prime}_0$ and $x^{\prime\prime\prime}_t$, according to Bayes' Theorem, we can obtain the probability distribution of $x^{\prime\prime\prime}_{t-1}$ as $p(x^{\prime\prime\prime}_{t-1}|x^{\prime\prime\prime}_{t},x^{\prime\prime\prime}_{0}){=}p(x^{\prime\prime\prime}_{t}|x^{\prime\prime\prime}_{t-1})\cdotp(x^{\prime\prime\prime}_{t-1}|x^{\prime\prime\prime}_{0})/p(x^{\prime\prime\prime}_{t}|x^{\prime\prime\prime}_{0})$. According to Eq.~\ref{eq: forward_ddpm} \cite{Diffusion_ho_2020}, $x^{\prime\prime\prime}_0$ can be approximate by 
\begin{equation}
    \begin{aligned}x^{\prime\prime\prime}_0=(1/{\sqrt{\overline{\alpha}_t}})\cdot(x^{\prime\prime\prime}_t-\sqrt{1-\overline{\alpha}_t}z_t),\end{aligned}
\end{equation}
where $z_t=\theta(x^{\prime\prime\prime}_t,t)$ is the estimation of the real noise in step $t$. Consequently, $x^{\prime\prime\prime}_{t-1}$ can be calculated by:
\begin{equation}
\label{eq: 14}
    \begin{array}{cc}
         x^{\prime\prime\prime}_{t-1}=\frac{1}{\sqrt{\alpha_t}}(x^{\prime\prime\prime}_t-\frac{1-\alpha_t}{\sqrt{1-\overline{\alpha}_t}}\theta(x^{\prime\prime\prime}_t,t))+\sigma_tz,\\
         \sigma_t^2=\frac{1-\overline{\alpha}_{t-1}}{1-\overline{\alpha}_t}\cdot\beta_t,
         \ z\sim \mathcal{N}(0,\mathbf{I}).
    \end{array}
\end{equation}
By iteratively applying this reverse step from $t = T$ down to $t = 1$ , we effectively recover the disturbed clean features by the forward process. While powerful, diffusion models are computationally expensive. To ensure efficiency, we set the total number of diffusion steps to a small value, $T=20$. This is considerably more lightweight than the 1000 steps used in methods like ZIP and the 140 steps used in the original SampDetox. As this stage adopts an existing methodology, we refer readers to the original SampDetox paper \cite{SampDetox_yang_2024} for more detailed theoretical guarantees of the DDPM-based purification process.

\begin{algorithm}[h]
\caption{\sysname}
\label{Algo: bdsec}
\renewcommand{\algorithmicrequire}{\textbf{Input:}}
\renewcommand{\algorithmicensure}{\textbf{Output:}}

    \begin{algorithmic}[1] 
        \REQUIRE  Backdoored inputs ($x$), the model set used for three stage ($\{f^{\prime}_{remove}, f^{\prime}_{recons},f^{\prime\prime},f^{\prime\prime\prime}\}$), the number of noise-adding and denoising steps in stage three ($T$);
	\ENSURE Purified inputs ($x_{pur}$); 
    
        // \textit{Stage I: Remove the high-visibility triggers}
        
        \STATE $m^{\prime} \leftarrow f^{\prime}_{remove}(x)$;
        
        \STATE $x^{\prime}_{recons} \leftarrow f^{\prime}_{recons}(x, m^{\prime})$;
        
        \STATE $x^\prime \leftarrow x \odot m^{\prime} + (1 - m^{\prime}) \odot x^{\prime}_{recons}$; // Eq.~\ref{eq: stage_1}.
        
        // \textit{Stage II: Remove the semi-visibility triggers}
        
        \STATE $x^{\prime\prime}_{recons} \leftarrow f^{\prime\prime}(x^\prime)$;

        \STATE $x^{\prime\prime}\leftarrow x^\prime - x^{\prime\prime}_{recons}$; // Eq.~\ref{eq: 11}
        
        // \textit{Stage III: Remove the low-visibility triggers}
        
        \STATE $x^{\prime\prime\prime}_{T}\leftarrow\sqrt{\overline{\alpha}_{T}}x^{\prime\prime} + \sqrt{1-\overline{\alpha}_{T}}z$; //$z\sim \mathcal{N}(0,\mathbf{I})$ 

        \FORALL{$t=T,\cdots,1$}
        \STATE $z\sim \mathcal{N}(0,\mathbf{I})$ if $t>1$, else $z = 0$;
        \STATE $\sigma_t^2\leftarrow({1-\overline{\alpha}_{t-1}})\cdot\beta_t/({1-\overline{\alpha}_t})$;
        \STATE $x^{\prime\prime\prime}_{t-1}=\frac{1}{\sqrt{\alpha_t}}(x^{\prime\prime\prime}_t-\frac{1-\alpha_t}{\sqrt{1-\overline{\alpha}_t}}\theta(x^{\prime\prime\prime}_t,t))+\sigma_tz$; // Eq.~\ref{eq: 14}
        \ENDFOR

        \STATE $x_{pur} = x^{\prime\prime\prime}_{t=1}$;
        
    \RETURN $x_{pur}$
    \end{algorithmic}
\end{algorithm}
\vspace{-5mm}    
\section{Experiments and Evaluations}
\label{sec: experiments}
\subsection{Experimental Setup}
\textbf{Datasets.} We conduct extensive evaluations across multiple datasets to demonstrate the effectiveness of \sysname. More concretely, we use three widely-used image classification datasets, i.e., CIFAR-10, CIFAR-100 \cite{cifar_krizhevsky_2009}, and ImageNette \cite{Imagenette_Howard_2019}. Descriptions of these datasets are as follows. \begin{enumerate*}
    \item \textbf{ImageNette} \cite{Imagenette_Howard_2019}: ImageNette is a small dataset extracted from the ImageNet \cite{imagenet_deng_2009}. It contains 9,469 training images and 3,925 test images in JPEG format, with non-uniform resolutions where both height and width are at least 160 pixels. In our experiments, we resize all images to a uniform resolution of $160\times160$ pixels using the Resize function from PyTorch.
    
    \item \textbf{CIFAR-10} \cite{cifar_krizhevsky_2009}: CIFAR-10 contains 60,000 $32\times32$ tiny images with 10 classes. In CIFAR-10, each class has 6,000 samples with 5,000 are training samples and 1,000 testing samples.
    
    \item \textbf{CIFAR-100} \cite{cifar_krizhevsky_2009}: CIFAR-100 contains 60,000 $32\times32$ tiny images with 100 classes. It is divided into 50,000 training images and 10,000 testing images, with 600 images per class.
\end{enumerate*}

\textbf{Networks.} We conduct our experiments on four deep learning models: PreActResNet-18, PreActResNet-34 \cite{preactresnet_he_2016}, MobileNet-V2 \cite{mobilenetv2_sandler_2018}, and Vision Transformer (ViT) \cite{vit_dosovitskiy_2020}. We implement the aforementioned models via the official code in BackdoorBench and maintain their default parameter settings. Note that unless otherwise stated, the default classification model is the PreActResNet-18.

\textbf{Metrics.} To evaluate the effectiveness of \sysname, we adopt three widely-used metrics according to SampDetox \cite{SampDetox_yang_2024}: Clean sample Accuracy (CA), Poisoned sample Accuracy (PA), and Attack Success Rate (ASR). CA measures the classification accuracy on benign samples after applying the defense, evaluating its impact on the model's original performance. A higher CA is desirable. PA refers to the accuracy on purified poisoned samples using their original labels. It measures the ability of a defense method to restore the correct features of poisoned inputs. A higher PA is better. ASR is the percentage of purified poisoned samples that are still misclassified as the target label. It directly evaluates the effectiveness of trigger removal, and a lower ASR is desirable. More details are provided in Appendix~\ref{sec: metric}.

\textbf{Attack Baselines.} We employ 11 SOTA backdoor attacks to evaluate the proposed method in our experiments, including four HVT-based attacks (BadNets \cite{BadNets_gu_2019}, InputAware \cite{IAD_nguyen_2020}, TrojanNN \cite{trojannn_liu_2018} and LC \cite{LC_turner_2019}), three SVT-based attacks (Blended \cite{Blended_chen_2017}, LF \cite{lf_zeng_2021} and SIG \cite{SIG_barni_2019}) and four LVT-based attacks (ISSBA \cite{issba_li_2021}, BPP \cite{BPP_wang_2022}, WaNet \cite{wanet_nguyen_2021}, and WaveAttack \cite{WaveAttack_xia_2024}). We implement the attacks using to the open-sourced backdoor attack toolboxes: BackdoorBench \cite{backdoorbench_wu_2022} and BackdoorBox \cite{backdoorbox_li_2023}. Key attack parameters are reported in Table~\ref{tab: Attack_Details}, and visualizations of the triggers can be found in the first row of Fig.~\ref{Fig: Purified_Results}.

\begin{table}[]
\caption{Attack Details}
\vspace{-2mm}
\label{tab: Attack_Details}
\begin{tabular}{cccc}
\toprule
Attacks    & Poison Rate & Trigger                            & Clean Label                \\ \midrule
BadNets    & 10\%        & 3*3 Grid                           & \textcolor{red}{\ding{55}} \\
InputAware & 10\%        & Dynamic                            & \textcolor{red}{\ding{55}} \\
TrojanNN   & 10\%        & Apple Logo                         & \textcolor{red}{\ding{55}} \\
LC         & 10\%        & 3*3 Grid *4                        & \textcolor{dark_green}{\ding{51}} \\
Blended    & 10\%      & HelloKitty, $\alpha=0.2$             & \textcolor{red}{\ding{55}} \\
LF         & 10\%        & Optimized                          & \textcolor{red}{\ding{55}} \\
SIG        & 10\%        & sinusoidal signal & \textcolor{dark_green}{\ding{51}} \\
ISSBA      & 10\%        & Dynamic                            & \textcolor{red}{\ding{55}} \\
BPP        & 10\% & Dynamic                            & \textcolor{red}{\ding{55}} \\
WaNet      & 10\%  & Dynamic                            & \textcolor{red}{\ding{55}} \\
WaveAttack & 5\%   & Dynamic                            & \textcolor{red}{\ding{55}} \\ \bottomrule
\end{tabular}
\vspace{-5mm}
\end{table}

\textbf{Defense Baselines.}
We compare \sysname with two state-of-the-art black-box backdoor defenses: ZIP \cite{ZIP_shi_2023} and SampDetox \cite{SampDetox_yang_2024}. We implement both baselines following their respective papers and official open-source code repositories \cite{ZIP_Github, SampDetox_Github}. Specifically, for ZIP, we utilized the guided-diffusion model from OpenAI \cite{ZIP_guided_diffusion}, as recommended by the original authors. For SampDetox, we utilized the pre-trained diffusion model provided by the authors for the CIFAR-10 task, while for other tasks, we trained new models following their official implementation. In terms of global and local purification time-steps, we follow their default setting, where $T_{global} = 20$ and $T_{local} = 120$.

\subsection{Purification Results}
\label{subsec: purification_results}

\begin{table*}[!]
\caption{Defense performance comparison. The values in parentheses indicates the percentage improvement (resp., degradation) of the best-performing baseline. For CA and PA, higher (\textcolor{red}{$\uparrow$}) is better; for ASR, lower (\textcolor{dark_green}{$\downarrow$}) is better. These results are visualized in Fig.~\ref{fig: comparison}.}

\label{tab: Performance comparison}
\resizebox{\linewidth}{!}{
\begin{tabular}{cc|ccc|ccc|ccc|ccc}

\toprule
\multicolumn{2}{c|}{\multirow{2}{*}{}}                        & \multicolumn{3}{c|}{No Defense} & \multicolumn{3}{c|}{ZIP} & \multicolumn{3}{c|}{SampDetox} & \multicolumn{3}{c}{Ours}                                                                                                                                        \\ \cmidrule{3-14} 
\multicolumn{2}{c|}{}                                         & CA$\uparrow$      & PA$\uparrow$      & ASR$\downarrow$     & CA$\uparrow$    & PA$\uparrow$    & ASR$\downarrow$  & CA$\uparrow$      & PA$\uparrow$      & ASR$\downarrow$    & CA$\uparrow$                                                 & PA$\uparrow$                                                 & ASR$\downarrow$                                                \\ \midrule
\multicolumn{1}{c|}{\multirow{11}{*}{CIFAR-10}}  & BadNets    & 91.33    & 4.66     & 95.03    & 74.71  & 17.28  & 81.10 & 86.58    & 71.90    & 14.12   & 87.13 (\textcolor{red}{$\uparrow$0.55})   & 84.73 (\textcolor{red}{$\uparrow$12.83})  & 2.38 (\textcolor{dark_green}{$\downarrow$11.74}) \\
\multicolumn{1}{c|}{}                            & InputAware & 90.67    & 1.65     & 98.25    & 75.89  & 58.60  & 14.10 & 84.18    & 48.57    & 49.49   & 86.51 (\textcolor{red}{$\uparrow$2.33})   & 82.83 (\textcolor{red}{$\uparrow$24.23})  & 3.47 (\textcolor{dark_green}{$\downarrow$10.63}) \\
\multicolumn{1}{c|}{}                            & TrojanNN   & 93.44    & 0.00     & 100.00   & 78.08  & 6.47   & 88.69 & 86.71    & 22.95    & 72.95   & 88.77 (\textcolor{red}{$\uparrow$2.06})   & 84.72 (\textcolor{red}{$\uparrow$61.77})  & 2.25 (\textcolor{dark_green}{$\downarrow$70.70}) \\
\multicolumn{1}{c|}{}                            & LC         & 91.79    & 0.04     & 99.95    & 79.73  & 6.17   & 93.33 & 85.59    & 13.10    & 86.50   & 87.63 (\textcolor{red}{$\uparrow$2.04})   & 87.28 (\textcolor{red}{$\uparrow$74.18})  & 1.42 (\textcolor{dark_green}{$\downarrow$85.08}) \\
\multicolumn{1}{c|}{}                            & Blended      & 93.47    & 0.07     & 99.92    & 77.97  & 30.70  & 29.30 & 86.59    & 38.10     & 54.34   & 88.52 (\textcolor{red}{$\uparrow$1.93})   & 74.69 (\textcolor{red}{$\uparrow$36.59})  & 4.01 (\textcolor{dark_green}{$\downarrow$25.29}) \\
\multicolumn{1}{c|}{}                            & LF         & 93.19    & 0.71     & 99.27    & 77.22  & 15.47  & 77.91 & 85.80    & 48.76    & 44.16   & 88.27 (\textcolor{red}{$\uparrow$2.47})   & 79.26 (\textcolor{red}{$\uparrow$30.50})  & 4.41 (\textcolor{dark_green}{$\downarrow$39.75}) \\
\multicolumn{1}{c|}{}                            & SIG        & 90.13    & 0.08     & 99.91    & 72.27  & 2.34   & 92.70 & 86.17    & 3.13     & 95.31   & 87.78 (\textcolor{red}{$\uparrow$1.61})   & 74.71 (\textcolor{red}{$\uparrow$71.58})  & 1.95 (\textcolor{dark_green}{$\downarrow$90.75}) \\
\multicolumn{1}{c|}{}                            & ISSBA      & 93.81    & 0.02     & 99.97    & 80.49  & 70.30  & 1.42  & 88.74    & 89.40    & 0.92    & 91.87 (\textcolor{red}{$\uparrow$3.13})   & 90.72 (\textcolor{red}{$\uparrow$1.32})   & 0.93 (\textcolor{red}{$\uparrow$0.01})    \\
\multicolumn{1}{c|}{}                            & BPP        & 90.69    & 0.21     & 99.77    & 79.43  & 70.53  & 3.39  & 86.86    & 81.09    & 1.76    & 87.82 (\textcolor{red}{$\uparrow$0.96})   & 82.77 (\textcolor{red}{$\uparrow$1.68})   & 1.27 (\textcolor{dark_green}{$\downarrow$0.49})  \\
\multicolumn{1}{c|}{}                            & WaNet      & 91.24    & 9.75     & 89.71    & 74.51  & 41.60  & 32.20 & 85.15    & 83.85    & 2.47    & 85.59 (\textcolor{red}{$\uparrow$0.44})   & 84.44 (\textcolor{red}{$\uparrow$0.59})   & 2.77 (\textcolor{red}{$\uparrow$0.30})    \\
\multicolumn{1}{c|}{}                            & WaveAttack & 92.33    & 8.77     & 90.84    & 78.94  & 67.81  & 3.12  & 82.99    & 72.91    & 2.69    & 84.57 (\textcolor{red}{$\uparrow$1.58})   & 80.69 (\textcolor{red}{$\uparrow$7.78})   & 2.76 (\textcolor{red}{$\uparrow$0.07})    \\ \midrule
\multicolumn{1}{c|}{\multirow{9}{*}{CIFAR-100}}  & BadNets    & 67.21    & 10.49    & 87.43    & 43.97  & 23.51  & 74.89 & 59.26    & 28.93    & 56.66   & 61.20 (\textcolor{red}{$\uparrow$1.94})   & 56.89 (\textcolor{red}{$\uparrow$27.96})  & 1.85 (\textcolor{dark_green}{$\downarrow$54.81}) \\
\multicolumn{1}{c|}{}                            & InputAware & 65.24    & 1.18     & 98.62    & 44.89  & 20.29  & 31.65 & 58.39    & 12.15    & 44.08   & 59.41 (\textcolor{red}{$\uparrow$1.02})   & 52.54 (\textcolor{red}{$\uparrow$32.25})  & 4.56 (\textcolor{dark_green}{$\downarrow$27.09}) \\
\multicolumn{1}{c|}{}                            & TrojanNN   & 69.90    & 0.01     & 99.98    & 46.51  & 21.01  & 77.35 & 59.10    & 21.52    & 50.02   & 59.81 (\textcolor{red}{$\uparrow$0.71})   & 53.39 (\textcolor{red}{$\uparrow$31.87})  & 0.13 (\textcolor{dark_green}{$\downarrow$49.89}) \\
\multicolumn{1}{c|}{}                            & Blended      & 70.48    & 5.42     & 93.66    & 45.11  & 16.28  & 17.72 & 57.33    & 16.19    & 61.66   & 60.41 (\textcolor{red}{$\uparrow$3.08})   & 49.28 (\textcolor{red}{$\uparrow$33.00})  & 8.87 (\textcolor{dark_green}{$\downarrow$8.85})  \\
\multicolumn{1}{c|}{}                            & LF         & 69.66    & 8.90     & 89.31    & 45.93  & 13.70  & 63.94 & 55.08    & 23.31    & 50.41   & 58.67 (\textcolor{red}{$\uparrow$3.59})   & 48.72 (\textcolor{red}{$\uparrow$25.41})  & 8.73 (\textcolor{dark_green}{$\downarrow$41.68}) \\
\multicolumn{1}{c|}{}                            & SIG        & 69.82    & 9.14     & 90.85    & 45.84  & 3.90   & 84.49 & 58.61    & 10.38    & 65.96   & 59.31 (\textcolor{red}{$\uparrow$0.70}) & 38.59 (\textcolor{red}{$\uparrow$28.21})  & 5.23 (\textcolor{dark_green}{$\downarrow$60.73}) \\
\multicolumn{1}{c|}{}                            & ISSBA      & 57.64    & 0.01     & 99.96    & 43.82  & 34.19  & 0.86  & 51.02    & 48.56    & 0.21    & 50.29 (\textcolor{dark_green}{$\downarrow$0.73}) & 50.63 (\textcolor{red}{$\uparrow$2.07})   & 0.44 (\textcolor{red}{$\uparrow$0.23})    \\
\multicolumn{1}{c|}{}                            & BPP        & 64.01    & 0.93     & 98.87    & 43.24  & 38.10  & 0.96  & 57.01    & 54.11    & 0.54    & 58.83 (\textcolor{red}{$\uparrow$1.82})   & 53.29 (\textcolor{dark_green}{$\downarrow$0.82}) & 0.35 (\textcolor{dark_green}{$\downarrow$0.19})  \\
\multicolumn{1}{c|}{}                            & WaNet      & 64.16    & 8.38     & 88.86    & 48.86  & 19.65  & 43.68 & 54.15    & 52.70    & 2.91    & 55.79 (\textcolor{red}{$\uparrow$1.64})   & 53.72 (\textcolor{red}{$\uparrow$1.02})   & 1.51 (\textcolor{dark_green}{$\downarrow$1.40})  \\ \midrule
\multicolumn{1}{c|}{\multirow{9}{*}{Imagenette}} & BadNets    & 89.60    & 1.52     & 98.47    & 71.39  & 18.11  & 80.09 & 83.98    & 20.55    & 79.16   & 87.98 (\textcolor{red}{$\uparrow$4.00})   & 84.28 (\textcolor{red}{$\uparrow$63.73})  & 1.24 (\textcolor{dark_green}{$\downarrow$77.92}) \\
\multicolumn{1}{c|}{}                            & InputAware & 82.29    & 4.63     & 94.68    & 71.03  & 10.61  & 87.11 & 79.02    & 16.08    & 59.51   & 79.29 (\textcolor{red}{$\uparrow$0.27})   & 76.42 (\textcolor{red}{$\uparrow$60.34})  & 0.48 (\textcolor{dark_green}{$\downarrow$59.03}) \\
\multicolumn{1}{c|}{}                            & TrojanNN   & 88.48    & 0.22     & 99.77    & 73.13  & 17.69  & 81.34 & 81.65    & 19.21    & 68.18   & 88.16 (\textcolor{red}{$\uparrow$6.51})   & 81.06 (\textcolor{red}{$\uparrow$61.85})  & 2.76 (\textcolor{dark_green}{$\downarrow$65.42}) \\
\multicolumn{1}{c|}{}                            & Blended      & 88.96    & 1.10     & 98.78    & 67.96  & 47.11  & 14.42 & 79.42    & 25.07    & 66.44   & 86.90 (\textcolor{red}{$\uparrow$7.48})   & 67.31 (\textcolor{red}{$\uparrow$20.20})  & 7.93 (\textcolor{dark_green}{$\downarrow$6.49})  \\
\multicolumn{1}{c|}{}                            & LF         & 88.22    & 3.44     & 96.21    & 69.47  & 1.92   & 97.61 & 81.89    & 2.31     & 96.80   & 87.12 (\textcolor{red}{$\uparrow$5.23})   & 81.03 (\textcolor{red}{$\uparrow$78.72})  & 1.24 (\textcolor{dark_green}{$\downarrow$95.56}) \\
\multicolumn{1}{c|}{}                            & SIG        & 88.63    & 3.33     & 96.55    & 72.17  & 6.73   & 91.30 & 85.31    & 7.36     & 92.41   & 87.35 (\textcolor{red}{$\uparrow$2.04})   & 74.22 (\textcolor{red}{$\uparrow$66.86})  & 8.76 (\textcolor{dark_green}{$\downarrow$82.54}) \\
\multicolumn{1}{c|}{}                            & ISSBA      & 82.83    & 0.39     & 99.57    & 73.10  & 70.19  & 5.96  & 80.16    & 80.04    & 1.59    & 82.11 (\textcolor{red}{$\uparrow$1.95})   & 79.05 (\textcolor{dark_green}{$\downarrow$0.99}) & 3.55 (\textcolor{red}{$\uparrow$1.96})    \\
\multicolumn{1}{c|}{}                            & BPP        & 83.31    & 0.99     & 98.86    & 77.64  & 76.53  & 1.29  & 81.33    & 80.19    & 0.43    & 81.84 (\textcolor{red}{$\uparrow$0.51})   & 80.70 (\textcolor{red}{$\uparrow$0.51})   & 0.07 (\textcolor{dark_green}{$\downarrow$0.36})  \\
\multicolumn{1}{c|}{}                            & WaNet      & 83.77    & 1.87     & 98.04    & 75.48  & 72.45  & 8.72  & 80.91    & 77.45    & 8.16    & 82.81 (\textcolor{red}{$\uparrow$1.90})   & 81.68 (\textcolor{red}{$\uparrow$4.23})   & 7.81 (\textcolor{dark_green}{$\downarrow$0.35})  \\ \bottomrule
\end{tabular}}
\vspace{-2mm}
\end{table*}

\begin{figure*}[!ht]
    \centering
    \includegraphics[width=1\linewidth]{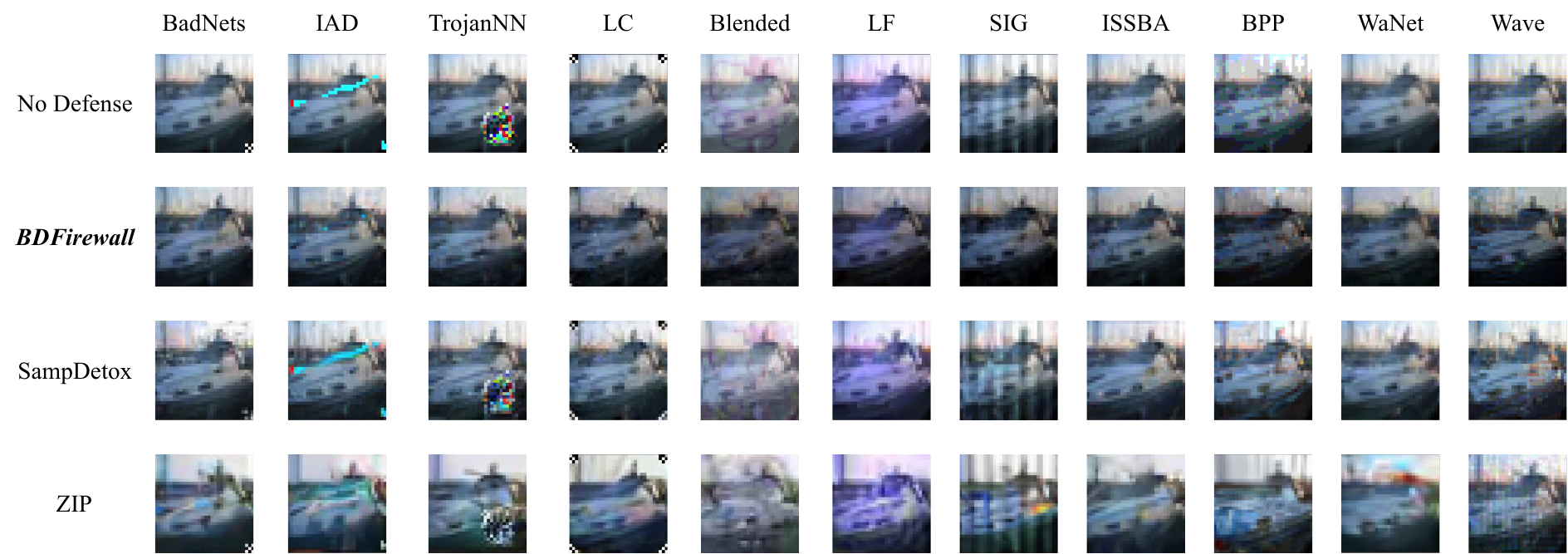}
    \caption{Visualization of purification results of different algorithms.}
    \label{Fig: Purified_Results}
    \vspace{-6mm}
\end{figure*}

Table~\ref{tab: Performance comparison} presents an extensive performance comparison between \sysname and two SOTA black-box backdoor defense baselines across three widely-used datasets. The results demonstrate that our proposed \sysname method achieves significantly improved purification effectiveness on backdoor samples, evidenced by a substantial reduction in Attack Success Rate (ASR). Concurrently, it incurs lower performance degradation on clean samples, as indicated by higher CA and PA scores. Specifically, on CIFAR-10, \sysname outperforms the best-performing baseline by 1.61\% in CA and 45.96\% in PA, while achieving a 56.71\% greater reduction in ASR. The corresponding results on CIFAR-100 are 1.84\%, 31.16\%, and 49.90\%, and on Imagenette are 4.26\%, 62.29\%, and 73.35\%. We attribute this superior performance to our carefully designed purification process. In Stage I, we leverage the significant semantic differences between high-visibility triggers and natural image features to identify clean regions within backdoor inputs and subsequently inpaint the corrupted areas to restore missing semantic information. This stage significantly mitigates the attack risk posed by residual trigger patterns. In the second stage, we reconstruct these obfuscated malicious patterns and then precisely remove them to yield a benign input. Combined, these two stages enable \sysname to effectively defend against HVT and SVT-based attacks, achieving an average ASR reduction of 59.99\% compared to the best-performing baseline across all scenarios.

In scenarios involving LVTs, \sysname demonstrates performance that is not only comparable to but, in most cases, superior to the best baseline method. The primary reason is that SampDetox, the strongest baseline in most scenarios, attempts to neutralize low-visibility triggers via global detoxification—a principle analogous to the third stage of our \sysname. However, as analyzed in Sec.~\ref{sec: intro}, SampDetox's reliance on inaccurate localization for its local detoxification process leads to a critical flaw: it fails to precisely isolate the trigger while erroneously corrupting clean, benign features. This flaw, however, acts as a double-edged sword. On one hand, the extensive diffusion steps can inadvertently disrupt low-visibility triggers, leading to a reduction in ASR. On the other hand, this same inaccuracy results in incomplete purification of high-visibility triggers and collateral damage to clean features, thereby reducing PA and CA. This trade-off explains why \sysname, with its precise, staged approach, consistently achieves superior overall performance. Overall, \sysname establishes a new state-of-the-art in black-box backdoor defense, demonstrating robust and stable performance across datasets of varying resolutions. 

Furthermore, we visualize the purification results for various backdoor attacks in Fig.~\ref{Fig: Purified_Results} which displays the original trigger-injected inputs in the first row, followed by the corresponding outputs after being processed by \sysname and the two baseline methods. The visualizations clearly illustrate the incomplete purification by the baseline methods, leaving residual artifacts that could still expose the model to backdoor attacks. In contrast, \sysname effectively eliminates diverse trigger patterns while maximally preserving the integrity of clean features. Note that, more detailed visualizations are provided in Appendix~\ref{sec: Details of Purification} (Fig.~\ref{fig: detail_comparison}), offering further visual evidence that corroborates the superior performance of \sysname.

\begin{table*}[!ht]
\centering
\caption{Performance of \sysname across different model architectures. Here, w/o CA (resp., w/o ASR) represents the clean sample accuracy (resp., ASR of backdoor samples) on compromised model without \sysname. In addition, we report the changes in CA and PA (resp., ASR) about w/o CA (resp., w/o ASR) after purification by \sysname in the brackets after the result. CA and PA decrease lower is better, but ASR drops larger is better. }
\label{tab: different_models}
\resizebox{0.9\linewidth}{!}{
\begin{tabular}{c|c|c|c|c|c|c|c|c}
\toprule
         Models                       &      Metrics        & BadNets                                             & InputAware                                           & Trojannn                                            & Blended                                              & SIG                                                  & BPP                                                 & WaNet                                               \\ \midrule
\multirow{5}{*}{PreActResNet18} & w/o CA & 91.33                                               & 90.67                                                & 93.44                                               & 93.47                                                & 90.13                                                & 90.69                                               & 91.24                                               \\
                                & w/o ASR      & 95.03                                               & 98.25                                                & 100.00                                              & 99.92                                                & 99.91                                                & 99.77                                               & 89.71                                               \\
                                & CA$\uparrow$          & 87.13 (\textcolor{dark_green}{$\downarrow$4.20}) & 86.51 (\textcolor{dark_green}{$\downarrow$4.16})  & 88.77 (\textcolor{dark_green}{$\downarrow$4.67}) & 88.52 (\textcolor{dark_green}{$\downarrow$4.95})  & 87.78 (\textcolor{dark_green}{$\downarrow$2.35})  & 87.82 (\textcolor{dark_green}{$\downarrow$2.87}) & 85.59 (\textcolor{dark_green}{$\downarrow$5.65}) \\
                                & PA$\uparrow$          & 84.73 (\textcolor{dark_green}{$\downarrow$6.60}) & 82.83 (\textcolor{dark_green}{$\downarrow$7.84})  & 84.72 (\textcolor{dark_green}{$\downarrow$8.72}) & 74.69 (\textcolor{dark_green}{$\downarrow$18.78}) & 74.71 (\textcolor{dark_green}{$\downarrow$15.42}) & 82.77 (\textcolor{dark_green}{$\downarrow$7.92}) & 84.44 (\textcolor{dark_green}{$\downarrow$6.80}) \\
                                & ASR$\downarrow$         & 2.38 (\textcolor{dark_green}{$\downarrow$92.65}) & 3.47 (\textcolor{dark_green}{$\downarrow$94.78})  & 2.25 (\textcolor{dark_green}{$\downarrow$97.75}) & 4.01 (\textcolor{dark_green}{$\downarrow$95.91})  & 1.95 (\textcolor{dark_green}{$\downarrow$97.96})  & 1.27 (\textcolor{dark_green}{$\downarrow$98.50}) & 2.77 (\textcolor{dark_green}{$\downarrow$86.94}) \\ \midrule
\multirow{5}{*}{PreActResNet34} & w/o CA & 92.50                                               & 91.05                                                & 93.73                                               & 93.73                                                & 93.76                                                & 91.20                                               & 90.46                                               \\
                                & w/o ASR      & 100.00                                              & 97.13                                                & 99.99                                               & 99.56                                                & 99.88                                                & 97.38                                               & 96.68                                               \\
                                & CA$\uparrow$          & 86.69 (\textcolor{dark_green}{$\downarrow$5.81}) & 84.88 (\textcolor{dark_green}{$\downarrow$6.17})  & 87.96 (\textcolor{dark_green}{$\downarrow$5.77}) & 90.12 (\textcolor{dark_green}{$\downarrow$3.61})  & 88.67 (\textcolor{dark_green}{$\downarrow$5.09})  & 86.70 (\textcolor{dark_green}{$\downarrow$4.50}) & 83.12 (\textcolor{dark_green}{$\downarrow$7.34}) \\
                                & PA$\uparrow$          & 84.11 (\textcolor{dark_green}{$\downarrow$8.39}) & 81.05 (\textcolor{dark_green}{$\downarrow$10.00}) & 85.29 (\textcolor{dark_green}{$\downarrow$8.44}) & 75.62 (\textcolor{dark_green}{$\downarrow$18.11}) & 72.33 (\textcolor{dark_green}{$\downarrow$21.43}) & 82.61 (\textcolor{dark_green}{$\downarrow$8.59}) & 81.54 (\textcolor{dark_green}{$\downarrow$8.92}) \\
                                & ASR$\downarrow$         & 2.21 (\textcolor{dark_green}{$\downarrow$97.79}) & 2.62 (\textcolor{dark_green}{$\downarrow$94.51})  & 1.83 (\textcolor{dark_green}{$\downarrow$98.16}) & 4.61 (\textcolor{dark_green}{$\downarrow$94.95})  & 1.96 (\textcolor{dark_green}{$\downarrow$97.92})  & 1.32 (\textcolor{dark_green}{$\downarrow$96.06}) & 5.97 (\textcolor{dark_green}{$\downarrow$90.71}) \\ \midrule
\multirow{5}{*}{MobileNet V2}   & w/o CA & 81.84                                               & 78.99                                                & 82.03                                               & 82.16                                                & 82.30                                                & 82.30                                               & 81.26                                               \\
                                & w/o ASR      & 99.99                                               & 95.81                                                & 99.88                                               & 97.29                                                & 98.94                                                & 99.20                                               & 91.15                                               \\
                                & CA$\uparrow$          & 77.97 (\textcolor{dark_green}{$\downarrow$3.87}) & 76.89 (\textcolor{dark_green}{$\downarrow$2.10})  & 78.73 (\textcolor{dark_green}{$\downarrow$3.30}) & 79.98 (\textcolor{dark_green}{$\downarrow$2.18})  & 78.31 (\textcolor{dark_green}{$\downarrow$3.99})  & 79.41 (\textcolor{dark_green}{$\downarrow$2.89}) & 78.12 (\textcolor{dark_green}{$\downarrow$3.14}) \\
                                & PA$\uparrow$          & 76.55 (\textcolor{dark_green}{$\downarrow$5.29}) & 75.28 (\textcolor{dark_green}{$\downarrow$3.71})  & 77.38 (\textcolor{dark_green}{$\downarrow$4.65}) & 68.93 (\textcolor{dark_green}{$\downarrow$13.23}) & 69.93 (\textcolor{dark_green}{$\downarrow$12.37}) & 74.90 (\textcolor{dark_green}{$\downarrow$7.40}) & 75.81 (\textcolor{dark_green}{$\downarrow$5.45}) \\
                                & ASR$\downarrow$         & 2.02 (\textcolor{dark_green}{$\downarrow$97.97}) & 1.08 (\textcolor{dark_green}{$\downarrow$94.73})  & 3.78 (\textcolor{dark_green}{$\downarrow$96.10}) & 4.96 (\textcolor{dark_green}{$\downarrow$92.33})  & 1.02 (\textcolor{dark_green}{$\downarrow$97.92})  & 2.25 (\textcolor{dark_green}{$\downarrow$96.95}) & 2.86 (\textcolor{dark_green}{$\downarrow$88.29}) \\ \midrule
\multirow{5}{*}{ViT}            & w/o CA & 95.85                                               & 93.62                                                & 95.89                                               & 96.40                                                & 96.06                                                & 97.00                                               & 94.70                                               \\
                                & w/o ASR      & 100.00                                              & 93.90                                                & 99.99                                               & 99.89                                                & 95.98                                                & 99.69                                               & 96.46                                               \\
                                & CA$\uparrow$          & 93.85 (\textcolor{dark_green}{$\downarrow$2.00}) & 88.01 (\textcolor{dark_green}{$\downarrow$5.61})  & 94.89 (\textcolor{dark_green}{$\downarrow$1.00}) & 94.80 (\textcolor{dark_green}{$\downarrow$1.60})  & 94.79 (\textcolor{dark_green}{$\downarrow$1.27})  & 94.00 (\textcolor{dark_green}{$\downarrow$3.00}) & 92.22 (\textcolor{dark_green}{$\downarrow$2.48}) \\
                                & PA$\uparrow$          & 91.41 (\textcolor{dark_green}{$\downarrow$4.44}) & 86.69 (\textcolor{dark_green}{$\downarrow$6.93})  & 90.60 (\textcolor{dark_green}{$\downarrow$5.29}) & 75.42 (\textcolor{dark_green}{$\downarrow$20.98}) & 75.86 (\textcolor{dark_green}{$\downarrow$20.20}) & 93.01 (\textcolor{dark_green}{$\downarrow$3.99}) & 90.11 (\textcolor{dark_green}{$\downarrow$4.59}) \\
                                & ASR$\downarrow$         & 0.73 (\textcolor{dark_green}{$\downarrow$99.27}) & 4.48 (\textcolor{dark_green}{$\downarrow$89.42})  & 1.88 (\textcolor{dark_green}{$\downarrow$98.11}) & 6.43 (\textcolor{dark_green}{$\downarrow$93.46})  & 6.89 (\textcolor{dark_green}{$\downarrow$89.09})  & 1.24 (\textcolor{dark_green}{$\downarrow$98.45}) & 1.60 (\textcolor{dark_green}{$\downarrow$94.86}) \\ \bottomrule
\end{tabular}}
\vspace{-5 mm }
\end{table*}
\subsection{Compatible to various models.}
In this section, We further evaluate the generalizability of \sysname across various model architectures. As \sysname operates in a black-box setting without access to model parameters or feedback (Sec.~\ref{sec: threat_model}), it is inherently model-agnostic. To validate this, we evaluated its defense capabilities against backdoor attacks on a diverse set of models, including PreAct-ResNet18, PreAct-ResNet34, MobileNetV2, and ViT. The results are reported in Tab.~\ref{tab: different_models}. The results show that \sysname consistently defends against backdoor attacks across all tested architectures, achieving an average ASR reduction of 95.06\% while exhibiting a minimal CA loss of only 3.77\%. Regarding PA, the average performance drop is under 10\%, a loss primarily attributable to two challenging SVT backdoor scenarios. After excluding these two attacks, the average PA loss drops to just 6.7\%. The primary challenge in these cases stems from the semi-visible triggers themselves disrupting benign features. This effect is exacerbated by the inherently lossy nature of the reconstruction process, collectively leading to the decrease in PA. Despite the minor PA degradation, the substantial ASR reduction confirms that \sysname effectively neutralizes these semi-visible triggers, successfully preventing the attacks. These experimental results, therefore, demonstrate that \sysname provides robust and consistent defense across diverse model architectures, confirming its model-agnostic nature.

\subsection{Robustness to Out-of-Distribution Surrogate Data}
By default, \sysname's internal purification models are trained on CIFAR-10, following the configuration of SampDetox. This section evaluates \sysname's performance under a more challenging constraint where its internal models are trained on a surrogate dataset (CIFAR-100) that is out-of-distribution (OOD) with respect to the target task (CIFAR-10). As shown in Table~\ref{tab: Surro_dataset}, despite challenging, the performance degradation of \sysname remains within an acceptable range. Across 11 attack types, the average CA decreased by only 5.31\%, PA by 4.88\%, while the ASR increased by a mere 1.14\%. The increase in ASR is most pronounced for the Blended, LF, and SIG attacks. This is because Stage II, which reconstructs trigger patterns by treating clean features as noise to be denoised, relies on the assumption that the purification model is familiar with the distribution of these "clean" features. The use of a surrogate dataset violates this assumption, leading to less precise de-clean features and consequently, a minor increase in the final ASR. Nevertheless, even under these challenging OOD conditions, \sysname still achieves a 24.49\% improvement in PA and a 29.23\% decrease in ASR compared to the best-performing baseline reported in Table~\ref{tab: Performance comparison}. This performance highlights the robustness of \sysname in defending against backdoor attacks in the MLaaS setting.

\begin{table}[!]
\caption{Robustness to Out-of-Distribution Surrogate Data.}
\label{tab: Surro_dataset}
\resizebox{\linewidth}{!}{
\begin{tabular}{c|ccc|ccc}
\toprule
\multirow{2}{*}{Attacks} & \multicolumn{3}{c|}{Default} & \multicolumn{3}{c}{CIFAR-100}                                                 \\ \cmidrule{2-7} 
                  & CA$\uparrow$         & PA$\uparrow$         & ASR$\downarrow$      & CA$\uparrow$                      & PA$\uparrow$                      & ASR$\downarrow$                    \\ \midrule
BadNets                  & 87.13        & 84.73        & 2.38            & 80.56   (\textcolor{dark_green}{$\downarrow$6.57}) & 79.34 (\textcolor{dark_green}{$\downarrow$5.39}) & 2.61 (\textcolor{red}{$\uparrow$0.23})     \\
InputAware               & 86.51        & 82.83        & 3.47            & 80.02   (\textcolor{dark_green}{$\downarrow$6.49}) & 77.11 (\textcolor{dark_green}{$\downarrow$5.72}) & 3.91 (\textcolor{red}{$\uparrow$0.44})     \\
TrojanNN                 & 88.77        & 84.72        & 2.25            & 82.75   (\textcolor{dark_green}{$\downarrow$6.02}) & 77.57 (\textcolor{dark_green}{$\downarrow$7.15}) & 2.61 (\textcolor{red}{$\uparrow$0.36})     \\
LC                       & 87.63        & 87.28        & 1.42            & 80.09   (\textcolor{dark_green}{$\downarrow$7.54}) & 82.51 (\textcolor{dark_green}{$\downarrow$4.77}) & 1.31 (\textcolor{dark_green}{$\downarrow$0.11}) \\
Blended                  & 88.52        & 74.69        & 4.01            & 83.82   (\textcolor{dark_green}{$\downarrow$4.70}) & 71.28 (\textcolor{dark_green}{$\downarrow$3.41}) & 5.23 (\textcolor{red}{$\uparrow$1.22})     \\
LF                       & 88.27        & 79.26        & 4.41            & 84.74   (\textcolor{dark_green}{$\downarrow$3.53}) & 72.30 (\textcolor{dark_green}{$\downarrow$6.96}) & 7.82 (\textcolor{red}{$\uparrow$3.41})     \\
SIG                      & 87.78        & 74.71        & 1.95            & 80.54   (\textcolor{dark_green}{$\downarrow$7.24}) & 70.19 (\textcolor{dark_green}{$\downarrow$4.52}) & 8.41 (\textcolor{red}{$\uparrow$6.46})     \\
ISSBA                    & 91.87        & 90.72        & 0.93            & 87.04   (\textcolor{dark_green}{$\downarrow$4.83}) & 87.51 (\textcolor{dark_green}{$\downarrow$3.21}) & 0.75 (\textcolor{dark_green}{$\downarrow$0.18}) \\
BPP                      & 87.82        & 82.77        & 1.27            & 82.50   (\textcolor{dark_green}{$\downarrow$5.32}) & 78.12 (\textcolor{dark_green}{$\downarrow$4.65}) & 1.02 (\textcolor{dark_green}{$\downarrow$0.25}) \\
WaNet                    & 85.59        & 84.44        & 2.77            & 82.04   (\textcolor{dark_green}{$\downarrow$3.55}) & 80.50 (\textcolor{dark_green}{$\downarrow$3.94}) & 1.23 (\textcolor{dark_green}{$\downarrow$1.54}) \\
WaveAttack               & 84.57        & 80.69        & 2.76            & 81.92   (\textcolor{dark_green}{$\downarrow$2.65}) & 76.78 (\textcolor{dark_green}{$\downarrow$3.91}) & 5.23 (\textcolor{red}{$\uparrow$2.47})  \\ \bottomrule
\end{tabular}}
\vspace{-5mm}
\end{table}

\subsection{Ablation Study}
In this section, we conduct ablation studies to validate the effectiveness of the three stages in \sysname and to examine the impact of different trigger removal sequences on the final results.

\textbf{\textit{Ablation Study on Defense Stages.}} \sysname comprises three distinct components designed to progressively remove high/semi-visibility/low-visibility triggers. To validate their individual contributions, we systematically deactivate each component and evaluate the corresponding defense performance, with the results presented in Table~\ref{tab: ab_different component_combination}. When the Stage I component is deactivated, the remaining two components fail to effectively defend against attacks with high-visibility triggers (i.e., BadNets, InputAware, and TrojanNN), causing the ASRs for these attacks to remain high. This observation holds for the other two components as well, indicating that each component is essential for addressing triggers within its designated scope. when Stage II is deactivated and Stage III is activate (i.e., Stage II \textcolor{red}{\ding{55}} and Stage III \textcolor{dark_green}{\ding{51}}), the Blended trigger is partially removed. This behavior is analogous to the global detoxification process in SampDetox. Unfortunately, this also demonstrates that global detoxification alone cannot completely remove features of semi-visibility triggers, resulting in residual triggers that can still activate the model's hidden backdoor. Another interesting observation occurs when Stage II is activate and Stage III is deactivated (Stage II \textcolor{dark_green}{\ding{51}} and Stage III \textcolor{red}{\ding{55}}): the ASRs for BPP and WaNet slightly decrease. This is because the prediction from $f^{\prime\prime}$ cannot be perfectly black (despite the loss term in Eq.~\ref{eq: clean_loss_in_stage_2}), which may inadvertently disrupt the subtle features of low-visibility triggers, rendering some of them ineffective. This experiment confirms that each stage of \sysname fulfills its intended role and that the stages do not exhibit negative interference with one another.

\begin{table*}[!]
\caption{Ablation Study of \sysname's Components. The symbols \textcolor{dark_green}{\ding{51}} and \textcolor{red}{\ding{55}} denote whether a component is activated or deactivated, respectively.}
\label{tab: ab_different component_combination}
\vspace{-2mm}
\resizebox{\linewidth}{!}{
\begin{tabular}{ccc|cc|cc|cc|cc|cc|cc|cc}
\toprule
\multicolumn{3}{c|}{}       & \multicolumn{2}{c|}{BadNets} & \multicolumn{2}{c|}{InputAware} & \multicolumn{2}{c|}{TrojanNN} & \multicolumn{2}{c|}{Blended} & \multicolumn{2}{c|}{SIG} & \multicolumn{2}{c|}{BPP} & \multicolumn{2}{c}{WaNet} \\ \midrule
Stage I & Stage II & Stage III & PA$\uparrow$      & ASR$\downarrow$     & PA$\uparrow$       & ASR$\downarrow$       & PA$\uparrow$      & ASR$\downarrow$      & PA$\uparrow$     & ASR$\downarrow$    & PA$\uparrow$    & ASR$\downarrow$   & PA$\uparrow$    & ASR$\downarrow$   & PA$\uparrow$    & ASR$\downarrow$    \\ \midrule
\textcolor{red}{\ding{55}}       & \textcolor{red}{\ding{55}}       & \textcolor{red}{\ding{55}}       & 4.66          & 95.03        & 1.65           & 98.25          & 0.00          & 100.00        & 0.07         & 99.92       & 0.71        & 99.27      & 0.21        & 99.77      & 9.75        & 89.71       \\
\textcolor{red}{\ding{55}}       & \textcolor{red}{\ding{55}}       & \textcolor{dark_green}{\ding{51}}       & 4.95          & 95.03        & 6.18           & 93.16          & 0.04          & 99.95         & 39.35        & 53.14       & 1.00        & 98.88      & 83.01       & 1.02       & 84.54       & 1.35        \\
\textcolor{red}{\ding{55}}       & \textcolor{dark_green}{\ding{51}}       & \textcolor{red}{\ding{55}}       & 5.08          & 94.91        & 12.20          & 79.67          & 0.00          & 100.00        & 70.21        & 26.14       & 71.84       & 14.94      & 20.83       & 76.04      & 35.87       & 59.73       \\
\textcolor{red}{\ding{55}}       & \textcolor{dark_green}{\ding{51}}       & \textcolor{dark_green}{\ding{51}}       & 6.24          & 92.98        & 18.93          & 63.43          & 0.21          & 99.73         & 74.43        & 4.03        & 74.53       & 2.71       & 81.22       & 1.07       & 84.80       & 2.56        \\
\textcolor{dark_green}{\ding{51}}       & \textcolor{red}{\ding{55}}       & \textcolor{red}{\ding{55}}       & 90.67         & 0.92         & 85.87          & 5.24           & 87.78         & 3.71          & 0.53         & 99.43       & 0.07        & 99.92      & 0.28        & 99.70      & 9.77        & 89.68       \\
\textcolor{dark_green}{\ding{51}}       & \textcolor{red}{\ding{55}}       & \textcolor{dark_green}{\ding{51}}       & 86.67         & 2.77         & 82.16          & 3.48           & 84.68         & 1.60          & 39.58        & 52.70       & 0.98        & 98.88      & 82.41       & 1.04       & 85.21       & 1.62        \\
\textcolor{dark_green}{\ding{51}}       & \textcolor{dark_green}{\ding{51}}       & \textcolor{red}{\ding{55}}       & 87.04         & 1.74         & 83.46          & 4.17           & 87.56         & 3.60          & 70.20        & 26.15       & 71.80       & 14.94      & 0.92        & 99.01      & 13.03       & 86.23       \\
\textcolor{dark_green}{\ding{51}}       & \textcolor{dark_green}{\ding{51}}       & \textcolor{dark_green}{\ding{51}}       & 84.73         & 2.38         & 82.83          & 3.47           & 84.72         & 2.25          & 74.69        & 4.01        & 74.71       & 1.95       & 82.77       & 1.27       & 84.44       & 2.77        \\ \bottomrule
\end{tabular}}
\vspace{-2mm}
\end{table*}

\textbf{\textit{Effectiveness of Different Purification Orders.}} We now evaluate the impact of different purification orders on the overall defense performance, with the results presented in Table~\ref{tab: ab_performance_different_sequences}. The relative order of Stage I and Stage II (e.g., comparing sequences {1,2,3}, {1,3,2}, and {2,1,3}) has only a minor effect on PA and ASR for high-visibility triggers, resulting in a 1.69\% PA loss and a 0.59\% ASR increase at worst. As discussed previously, the inaccurate predictions from $f^{\prime\prime}$ can slightly corrupt the semantic cues used to locate clean features. However, the placement of Stage III is far more critical. Executing Stage III first leads to a substantial degradation in defense performance for both high-visibility and semi-visibility triggers, causing a 15.43\% PA loss and a 13.59\% ASR increase for the former, and an 18.94\% PA loss and a 23.05\% ASR increase for the latter. This side effect also propagates, leading to inaccurate predictions in Stage I and II and consequently a slight PA decrease for LVTs. The reason is that the diffusion process in Stage III disrupts the evidence required by Stage I. Specifically, the diffusion smooths the sharp semantic margins that Stage I relies on to identify segmentation shifts, rendering it ineffective. Furthermore, Stage III intensifies the mixing of clean and backdoor features within the image, making them more difficult to separate. This problem is reflected in the results for SVT-based attacks like Blended and SIG, where the ASR increases by 23.05\% and the PA decreases by 18.94\%. The results of this ablation study thus confirm our analysis in Sec.~\ref{sec: details design} regarding the critical importance of the proposed purification sequence.

\begin{table*}[]
\centering
\caption{Impact of Purification Order on Defense Performance. Performance metrics are compared against the baseline sequence (Stage I $\rightarrow$ Stage II $\rightarrow$ Stage III). Colored arrows indicate the performance change relative to the baseline: \textcolor{red}{$\uparrow$} signifies an increase in the metric's value, while \textcolor{dark_green}{$\downarrow$} signifies a decrease.}
\label{tab: ab_performance_different_sequences}
\resizebox{.75\linewidth}{!}{
\begin{tabular}{c|c|c|c|c|c|c|c}
\toprule
                            & Stage           & 1, 2, 3 & 1,3,2                                                                 & 2,1,3                                                                & 2,3,1                                                                 & 3,1,2                                                                 & 3,2,1                                                                 \\ \midrule
\multirow{2}{*}{BadNets}    & PA$\uparrow$    & 84.73   & 84.05 (\textcolor{dark_green}{$\downarrow$0.68})  & 83.21 (\textcolor{dark_green}{$\downarrow$1.52}) & 55.98 (\textcolor{dark_green}{$\downarrow$28.75}) & 62.84 (\textcolor{dark_green}{$\downarrow$21.89}) & 60.34 (\textcolor{dark_green}{$\downarrow$24.39}) \\
                            & ASR$\downarrow$ & 2.38    & 2.77 (\textcolor{red}{$\uparrow$0.39})             & 0.65 (\textcolor{dark_green}{$\downarrow$1.73})  & 33.28 (\textcolor{red}{$\uparrow$30.90})           & 24.95 (\textcolor{red}{$\uparrow$22.57})           & 28.34 (\textcolor{red}{$\uparrow$25.96})           \\ \midrule
\multirow{2}{*}{InputAware} & PA$\uparrow$    & 82.83   & 81.77 (\textcolor{dark_green}{$\downarrow$1.06})  & 81.04 (\textcolor{dark_green}{$\downarrow$1.79}) & 74.93 (\textcolor{dark_green}{$\downarrow$7.90})  & 75.32 (\textcolor{dark_green}{$\downarrow$7.51})  & 74.73 (\textcolor{dark_green}{$\downarrow$8.10})  \\
                            & ASR$\downarrow$ & 3.47    & 2.31 (\textcolor{dark_green}{$\downarrow$1.16})   & 3.75 (\textcolor{red}{$\uparrow$0.28})            & 8.78 (\textcolor{red}{$\uparrow$5.31})             & 8.46 (\textcolor{red}{$\uparrow$4.99})             & 9.12 (\textcolor{red}{$\uparrow$5.65})             \\ \midrule
\multirow{2}{*}{TrojanNN}   & PA$\uparrow$    & 84.72   & 82.53 (\textcolor{dark_green}{$\downarrow$2.19})  & 81.84 (\textcolor{dark_green}{$\downarrow$2.88}) & 70.80 (\textcolor{dark_green}{$\downarrow$13.92}) & 71.88 (\textcolor{dark_green}{$\downarrow$12.84}) & 71.18 (\textcolor{dark_green}{$\downarrow$13.54}) \\
                            & ASR$\downarrow$ & 2.25    & 1.63 (\textcolor{dark_green}{$\downarrow$0.62})   & 1.56 (\textcolor{dark_green}{$\downarrow$0.69})  & 11.68 (\textcolor{red}{$\uparrow$9.43})            & 10.66 (\textcolor{red}{$\uparrow$8.41})            & 11.35 (\textcolor{red}{$\uparrow$9.10})            \\ \midrule
\multirow{2}{*}{Blended}    & PA$\uparrow$    & 74.69   & 52.54 (\textcolor{dark_green}{$\downarrow$22.15}) & 74.08 (\textcolor{dark_green}{$\downarrow$0.61}) & 73.16 (\textcolor{dark_green}{$\downarrow$1.53})  & 54.74 (\textcolor{dark_green}{$\downarrow$19.95}) & 54.35 (\textcolor{dark_green}{$\downarrow$20.34}) \\
                            & ASR$\downarrow$ & 4.01    & 26.32 (\textcolor{red}{$\uparrow$22.31})           & 4.00 (\textcolor{dark_green}{$\downarrow$0.01})  & 3.29 (\textcolor{dark_green}{$\downarrow$0.72})   & 26.38 (\textcolor{red}{$\uparrow$22.37})           & 26.21 (\textcolor{red}{$\uparrow$22.20})           \\ \midrule
\multirow{2}{*}{SIG}        & PA$\uparrow$    & 74.71   & 57.74 (\textcolor{dark_green}{$\downarrow$16.97}) & 75.33 (\textcolor{red}{$\uparrow$0.62})           & 74.31 (\textcolor{dark_green}{$\downarrow$0.40})  & 57.55 (\textcolor{dark_green}{$\downarrow$17.16}) & 57.64 (\textcolor{dark_green}{$\downarrow$17.07}) \\
                            & ASR$\downarrow$ & 1.95    & 26.00 (\textcolor{red}{$\uparrow$24.05})           & 0.33 (\textcolor{dark_green}{$\downarrow$1.62})  & 4.84 (\textcolor{red}{$\uparrow$2.89})             & 25.93 (\textcolor{red}{$\uparrow$23.98})           & 25.35 (\textcolor{red}{$\uparrow$23.40})           \\ \midrule
\multirow{2}{*}{BPP}        & PA$\uparrow$    & 82.77   & 81.08 (\textcolor{dark_green}{$\downarrow$1.69})  & 79.68 (\textcolor{dark_green}{$\downarrow$3.09}) & 80.58 (\textcolor{dark_green}{$\downarrow$2.19})  & 81.28 (\textcolor{dark_green}{$\downarrow$1.49})  & 81.66 (\textcolor{dark_green}{$\downarrow$1.11})  \\
                            & ASR$\downarrow$ & 1.27    & 1.17 (\textcolor{dark_green}{$\downarrow$0.10})   & 1.13 (\textcolor{dark_green}{$\downarrow$0.14})  & 1.02 (\textcolor{dark_green}{$\downarrow$0.25})   & 1.04 (\textcolor{dark_green}{$\downarrow$0.23})   & 0.95 (\textcolor{dark_green}{$\downarrow$0.32})   \\ \midrule
\multirow{2}{*}{WaNet}      & PA$\uparrow$    & 84.44   & 84.18 (\textcolor{dark_green}{$\downarrow$0.26})  & 82.85 (\textcolor{dark_green}{$\downarrow$1.59}) & 83.42 (\textcolor{dark_green}{$\downarrow$1.02})  & 84.11 (\textcolor{dark_green}{$\downarrow$0.33})  & 84.30 (\textcolor{dark_green}{$\downarrow$0.14})  \\
                            & ASR$\downarrow$ & 2.77    & 1.47 (\textcolor{dark_green}{$\downarrow$1.30})   & 1.48 (\textcolor{dark_green}{$\downarrow$1.29})  & 1.45 (\textcolor{dark_green}{$\downarrow$1.32})   & 1.43 (\textcolor{dark_green}{$\downarrow$1.34})   & 1.10 (\textcolor{dark_green}{$\downarrow$1.67})   \\ \bottomrule
\end{tabular}}
\vspace{-6mm}
\end{table*}

\begin{table}[]
\centering
    \caption{Inference Time (s)}
    \label{tab: inference_time}
\vspace{-3mm}
\begin{tabular}{c|c|c|c}
\toprule
                   & ZIP   & SampDetox & \sysname     \\ \midrule
Sample-Level           & 7.49  & 0.40    & 0.12 (\textcolor{red}{$\uparrow\sim$3$\times$})  \\ \midrule
Batch-Level (64)        & 39.68 & 118.77    & 0.65 (\textcolor{red}{$\uparrow\sim$61$\times$})  \\ \midrule
Batch-Level (128)      & 89.59     & 238.48    & 0.81 (\textcolor{red}{$\uparrow\sim$111$\times$})   \\ \bottomrule
\end{tabular}
\vspace{-6mm}
\end{table}

\subsection{Inference Time}
In this section, we evaluate the inference time of ZIP, SampDetox, and \sysname, which is a critical factor for deploying black-box defense methods in practical applications. Specifically, we report both sample-level and batch-level inference times in Table~\ref{tab: inference_time}, where sample-level time is defined as the duration to process a single sample, and batch-level time is the duration to process an entire batch of samples. The results show that \sysname purifies suspicious samples significantly faster, achieving up to a 111$\times$ speedup compared to the best-performing baseline, particularly as the batch size increases. This efficiency is primarily attributed to Stage I and Stage II of \sysname, which effectively remove suspected triggers with one-step inference. In contrast, both baseline methods require multiple reverse diffusion processes to eliminate backdoor triggers. For instance, removing a high-visibility trigger requires SampDetox nearly 120 diffusion steps, even under ideal conditions. Therefore, DDPM-based methods inevitably cause inference delays due to their high computational overhead, making it challenging for users to obtain prediction results in a timely manner. We note that while ZIP suffers from a tremendous inference delay for a single sample, it performs well at the batch level. This is because we follow its official implementation, which batches 64 samples into a single 256×256 input for the diffusion model, thereby saving considerable time. Compared with SampDetox, \sysname achieves an even greater speedup of up to 294$\times$. This highlights the superior expeditiousness of \sysname across various inference scales relative to other DDPM-based methods. Additionally, we discuss the use of DDIM to further accelerate inference for \sysname in Sec.~\ref{subsec: ddim}.

\begin{table*}[!ht]
\caption{Performance with DDIM. The result in brackets is a comparison between variations and \sysname.}
\label{tab: ddim}
\centering
\resizebox{.85\linewidth}{!}{
\begin{tabular}{@{}c|c|c|c|c|c|c|c|c@{}}
\toprule
                            &                  & BadNets                                          & InputAware                                       & TrojanNN                                         & Blend                                            & SIG                                              & BPP                                              & WaNet                                            \\ \midrule
\multirow{3}{*}{\sysname}   & Time$\downarrow$ & 0.65                                             & 0.65                                             & 0.65                                             & 0.65                                             & 0.65                                             & 0.65                                             & 0.64                                             \\
                            & PA$\uparrow$     & 84.73                                            & 82.83                                            & 84.72                                            & 74.69                                            & 74.71                                            & 82.77                                            & 84.44                                            \\
                            & ASR$\downarrow$  & 2.38                                             & 3.47                                             & 2.25                                             & 4.01                                             & 1.95                                             & 1.27                                             & 2.77                                             \\ \midrule
\multirow{3}{*}{\sysname-5} & Time$\downarrow$ & 0.29(\textcolor{blue}{$\sim$2.24$\times$})       & 0.29(\textcolor{blue}{$\sim$2.24$\times$})       & 0.29(\textcolor{blue}{$\sim$2.24$\times$})       & 0.29(\textcolor{blue}{$\sim$2.27$\times$})       & 0.29(\textcolor{blue}{$\sim$2.24$\times$})       & 0.29(\textcolor{blue}{$\sim$2.27$\times$})       & 0.29(\textcolor{blue}{$\sim$2.24$\times$})       \\
                            & PA$\uparrow$     & 81.78(\textcolor{dark_green}{$\downarrow$2.95})  & 79.87(\textcolor{dark_green}{$\downarrow$2.96})  & 81.02(\textcolor{dark_green}{$\downarrow$3.70})  & 61.83(\textcolor{dark_green}{$\downarrow$12.86}) & 55.08(\textcolor{dark_green}{$\downarrow$19.63}) & 72.68(\textcolor{dark_green}{$\downarrow$10.09}) & 76.15(\textcolor{dark_green}{$\downarrow$8.29})  \\
                            & ASR$\downarrow$  & 0.53(\textcolor{dark_green}{$\downarrow$1.85})   & 2.32(\textcolor{dark_green}{$\downarrow$1.15})   & 1.05(\textcolor{dark_green}{$\downarrow$1.20})   & 8.12(\textcolor{red}{$\uparrow$4.11})            & 6.42(\textcolor{red}{$\uparrow$4.47})            & 1.23(\textcolor{dark_green}{$\downarrow$0.04})   & 1.04(\textcolor{dark_green}{$\downarrow$1.73})   \\ \midrule
\multirow{3}{*}{\sysname-1} & Time$\downarrow$ & 0.08(\textcolor{blue}{$\sim$8.13$\times$})       & 0.08(\textcolor{blue}{$\sim$8.13$\times$})       & 0.07(\textcolor{blue}{$\sim$8.90$\times$})       & 0.08(\textcolor{blue}{$\sim$8.13$\times$})       & 0.08(\textcolor{blue}{$\sim$8.44$\times$})       & 0.08(\textcolor{blue}{$\sim$8.55$\times$})       & 0.08(\textcolor{blue}{$\sim$8.31$\times$})       \\
                            & PA$\uparrow$     & 44.15(\textcolor{dark_green}{$\downarrow$40.58}) & 47.51(\textcolor{dark_green}{$\downarrow$35.32}) & 41.70(\textcolor{dark_green}{$\downarrow$43.02}) & 19.08(\textcolor{dark_green}{$\downarrow$55.61}) & 28.67(\textcolor{dark_green}{$\downarrow$46.04}) & 56.30(\textcolor{dark_green}{$\downarrow$26.47}) & 46.97(\textcolor{dark_green}{$\downarrow$37.47}) \\
                            & ASR$\downarrow$  & 0.25(\textcolor{dark_green}{$\downarrow$2.13})   & 2.50(\textcolor{dark_green}{$\downarrow$0.97})   & 0.62(\textcolor{dark_green}{$\downarrow$1.63})   & 18.05(\textcolor{red}{$\uparrow$14.04})          & 18.21(\textcolor{red}{$\uparrow$16.26})          & 18.22(\textcolor{red}{$\uparrow$16.95})          & 0.21(\textcolor{dark_green}{$\downarrow$2.56})   \\ \midrule
\multirow{3}{*}{\sysname-0} & Time$\downarrow$ & 0.02(\textcolor{blue}{$\sim$32.50$\times$})      & 0.02(\textcolor{blue}{$\sim$30.95$\times$})      & 0.02(\textcolor{blue}{$\sim$30.95$\times$})      & 0.02(\textcolor{blue}{$\sim$30.95$\times$})      & 0.02(\textcolor{blue}{$\sim$30.95$\times$})      & 0.02(\textcolor{blue}{$\sim$30.95$\times$})      & 0.02(\textcolor{blue}{$\sim$32.00$\times$})      \\
                            & PA$\uparrow$     & 29.93(\textcolor{dark_green}{$\downarrow$54.80}) & 37.53(\textcolor{dark_green}{$\downarrow$45.30}) & 27.93(\textcolor{dark_green}{$\downarrow$56.79}) & 17.28(\textcolor{dark_green}{$\downarrow$57.41}) & 15.47(\textcolor{dark_green}{$\downarrow$59.24}) & 50.31(\textcolor{dark_green}{$\downarrow$32.46}) & 45.94(\textcolor{dark_green}{$\downarrow$38.50}) \\
                            & ASR$\downarrow$  & 0.17(\textcolor{dark_green}{$\downarrow$2.21})   & 0.93(\textcolor{dark_green}{$\downarrow$2.54})   & 0.27(\textcolor{dark_green}{$\downarrow$1.98})   & 28.97(\textcolor{red}{$\uparrow$24.96})          & 53.26(\textcolor{red}{$\uparrow$51.31})          & 29.73(\textcolor{red}{$\uparrow$28.46})          & 0.10(\textcolor{dark_green}{$\downarrow$2.67})   \\ \bottomrule
\end{tabular}}
\vspace{-7 mm}
\end{table*}

\section{Discussion}
\label{sec: discussion}
This paper presented \sysname, a novel approach for black-box backdoor defense in MLaaS environments. In this section, we discuss its primary limitations and the crucial balance between performance and computational overhead.

\subsection{Limitations of \sysname}
In the previous sections, we conducted extensive experiments across various datasets and models, demonstrating that the proposed \sysname can effectively and efficiently defend against backdoor attacks in black-box settings. Despite these successes, \sysname still has limitations in handling certain types of triggers. The core mechanism of \sysname, for defending against both HVT and SVT, relies on a key observation: the semantic differentiability between clean features and trigger features. Consequently, when triggers are semantically inseparable from the clean features, as seen in so-called semantic backdoor attacks, \sysname is unable to remove them effectively. 

\begin{figure}
    \centering
    \includegraphics[width=0.8\linewidth]{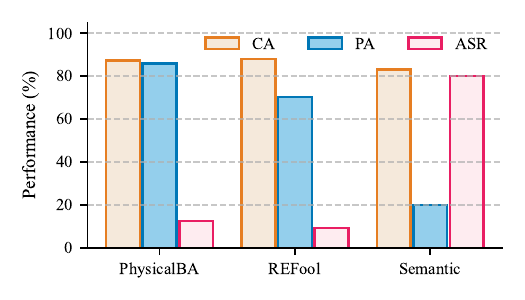}
    \caption{Physical Backdoor and Semantic Backdoor}
    \label{fig: limitation}
    \vspace{-8 mm}
\end{figure}

To investigate this limitation, we conducted experiments on three representative attacks: PhysicalBA \cite{PhysicalBA_li_2021}, REFool \cite{REFool_liu_2020}, and the Semantic Backdoor attack \cite{semantic_bagdasaryan_2020}. Figure~\ref{fig: limitation} presents the defense performance of \sysname against the aforementioned attacks. Surprisingly, \sysname defended against PhysicalBA and REFool with satisfactory PAs and ASRs, a result contrary to our initial hypothesis. Upon re-examining the backdoor generation process for PhysicalBA and REFool, we found that although they are categorized as semantic attacks, the triggers they apply are not intrinsic to the original clean features. This means the backdoor triggers still exhibit semantic differences from the clean features of the input images. Such semantic differences allow \sysname to successfully identify and neutralize them.

However, \sysname fails to defend against the Semantic Backdoor attack \cite{semantic_bagdasaryan_2020}, which leverages green cars as poisoned samples to induce the backdoor. In this attack, the backdoored model is trained to misclassify any green car to an adversary-specified label. Since the backdoor trigger (the color green) is an intrinsic feature of the benign samples (cars), our key observation regarding semantic differences no longer holds, causing \sysname to fail. Fortunately, the applicability of such attacks is limited, as they require a specific subclass of samples for poisoning, unlike more general attacks that can implant triggers into arbitrary images. We therefore conclude that while \sysname has this inherent limitation, its practical impact is confined to a narrow class of attacks, leaving the versatility of \sysname intact for most real-world scenarios.

\subsection{Balancing timeliness and effectiveness}
\label{subsec: ddim}
Diffusion models, with their outstanding performance in high-quality image generation and restoration, are becoming a core approach in generative artificial intelligence. Consequently, their powerful image restoration capabilities have been leveraged in recent research to purify backdoored samples, enabling backdoor defense in black-box scenarios. In this paper, we adopt this line of work, defending against low-visibility triggers by first adding lightweight noise to suspicious inputs and then applying a DDPM-based method to denoise them. Although effective, a critical problem of DDPM-based methods still lingers: \textit{their significant computational overhead.} As dictated by Eq.~\ref{eq: 14}, the prediction at step $t-1$ depends on the state at step $t$, meaning DDPM-based methods must execute the full sequence of predefined time steps iteratively, resulting in substantial computational costs. Although our method employs only 20 time steps and achieves an inference acceleration of up to 294$\times$ compared to SampDetox, a state-of-the-art DDPM-based method, its latency is still significant—approximately 30$\times$ slower than vanilla inference. This raises a critical question: \textit{how can we further accelerate the purification process?}

One promising solution is to apply Denoising Diffusion Implicit Models (DDIM) \cite{DDIM_song_2021}, which accelerate image generation through a non-Markovian process. To explore this trade-off, we introduce DDIM into our framework and propose three variants: \sysname-5, \sysname-1, and \sysname-0. Here, the number in \sysname-5 and \sysname-1 represents the total number of DDIM steps, while \sysname-0 represents a baseline where we only add noise to suspicious inputs without any denoising. As reported in Tab.~\ref{tab: ddim}, the results reveal that while DDIM significantly accelerates the purification process, it introduces a step-dependent performance degradation, particularly in PA, which shows a clear downward trend as the number of steps decreases. Thus, how to balance the trade-off between purification efficiency and defense effectiveness remains a critical open question. We believe this direction warrants further investigation from the research community.

\section{Conclusion}
\label{sec: conclusion}

This paper introduced \sysname, a progressive black-box backdoor defense framework based on a refined categorization of triggers: Fully-Visible Triggers (FVT), Semi-Visible Triggers (SVT), and Low-Visibility Triggers (LVT) according to their impact to the patch area. Specifically, we first capitalize the clear semantic difference of HVT attaching to identify and reconstruct trigger regions, thereby neutralizing the threat while recovering the original content. Then, we model SVT patched inputs as distinct trigger patterns corrupted by clean features and employ a denoising process to precisely reconstruct and subsequently eliminate the trigger. In order to further eliminate the LVT, we leverage their inherent fragility to noise by perturbing suspicious inputs to destroy them and then applying DDPM to purify them within a few diffusion steps. Extensive experiments demonstrate that \sysname significantly outperforms state-of-the-art black-box defenses. Compared to the best performance of them, \sysname achieves an average of 33.25\% ASR drop, improving poisoned sample accuracy by 29.64\%, and accelerating inference time by up to 111$\times$.

For future work, we identify two critical research directions. First, developing defenses against semantic backdoors, where triggers are intrinsically part of the clean features, remains a major challenge. Second, further optimizing the trade-off between purification effectiveness and computational latency is crucial for real-world deployment, potentially by exploring non-iterative purification models.

\bibliographystyle{IEEEtran}
\bibliography{sample}

\appendices
\section{Details of Metrics}
\label{sec: metric}
\begin{itemize}
    \item \textbf{Clean sample Accuracy (CA)} measures the accuracy of the backdoored model on clean samples after they have been processed by our defense method. A higher CA indicates that our defense preserves the model's utility on benign inputs. It is calculated as:
    \begin{equation}
    \text{CA} = \frac{1}{|D_{cln}|} \sum_{(x_i, y_i) \in D_{cln}} \mathbb{I}(f(P(x_i)) = y_i),
    \end{equation}
    where $D_{cln}$ is the set of clean test samples, $(x_i, y_i)$ is a sample-label pair from $D_{cln}$, $P(\cdot)$ is the purification function of \sysname, $f(\cdot)$ is the backdoored model, and $\mathbb{I}(\cdot)$ is the indicator function.

    \item \textbf{Poisoned sample Accuracy (PA)} refers to the accuracy of the backdoored model in predicting the original, ground-truth labels of poisoned samples after they have been purified by \sysname. This metric helps to evaluate whether the defense can successfully remove the trigger effect and restore the sample's original features. It is defined as:
    \begin{equation}
    \text{PA} = \frac{1}{|D_{bd}|} \sum_{(x_i, y_i) \in D_{bd}} \mathbb{I}(f(P(x_i)) = y_i),
    \end{equation}
    where $D_{bd}$ is the set of poisoned test samples and $y_i$ is the original ground-truth label for the sample $x_i$.

    \item \textbf{Attack Success Rate (ASR)} is the percentage of purified poisoned samples that are still misclassified as the attacker's target label by the backdoored model. It directly measures the effectiveness of our defense in mitigating the backdoor attack. A lower ASR is desirable. The formula is:
    \begin{equation}
    \text{ASR} = \frac{1}{|D_{bd}|} \sum_{x_i \in D_{bd}} \mathbb{I}(f(P(x_i)) = y_{target}),
    \end{equation}
    where $y_{target}$ is the attacker-specified target label.
\end{itemize}

\section{Implementation Details of \sysname}

\textbf{\textit{Model Architecture.}} In Stages I and II, we leverage three carefully designed U-Net models $f^{\prime}_{remove}$, $f^{\prime}_{recons}$ and $f^{\prime\prime}$ to defend against the high-visibility triggers and semi-visibility triggers in black-box settings. In terms of $f^{\prime}_{remove}$, it contains an encoder (four convolutions and two max-pooling layers for feature extraction), a decoder (two transposed convolutions and two regular convolutions for mask prediction) and a self-calibration module from SLBR \cite{SLBR_liang_2021} to guide the final prediction. For $f^{\prime}_{recons}$, it contains an encoder (three convolutions layers for feature extraction), one decoder (two transposed convolutions and one regular convolution for reconstructing the clean image). For $f^{\prime\prime}$, it contains a encoder with three double-convolution layers and three max-pooling layers for feature extraction, a bottleneck with a double-convolution layer for processing the middle feature and three transposed convolution layers for image generation. In Stage III, we employ a DDPM-based diffusion model, adapted from SampDetox, to purify inputs against LVTs.

\textbf{\textit{Training Details.}} We train $f^{\prime}_{remove}$, $f^{\prime}_{recons}$ and $f^{\prime\prime}$ for 200 epochs using the AdamW optimizer \cite{AdamW_loshchilov_2019} with a learning rate of 0.001. In terms of the diffusion model used in Stage III, we directly use the pre-trained model provided by SampDetox for the CIFAR-10 task, and utilize the training code of SampDetox to train new diffusion models for CIFAR-100 and Imagenette.

\textbf{\textit{Surrogate Triggers.}} Since we cannot access the training data of the model provided by the service provider, we have no way of obtaining information about the triggers. For this reason, we use different surrogate triggers in Stage I and Stage II, according to our observations of different types of triggers. In Stage I, we use randomly generated triangles, squares, and circles as shapes and fill them with various patterns, such as mosaics, random noise, and solid colors to manually create clear semantic distinctions. In Stage II, we combine shapes with different colors and randomly generated backgrounds for them to generate surrogate triggers whose size matches that of the inputs, and then randomly overlay them on the original samples with random transparency ($10\%\sim 40\%$). This process trains $f^{\prime\prime}$ to treat the benign image content as "noise" and effectively separate it from the underlying trigger pattern.

\begin{table}
\caption{Performances on Different Trigger Patterns. The number after represents the gap compared to the results in Table~\ref{tab: Performance comparison}. For CA and PA, higher (i.e., \textcolor{red}{$\uparrow$}) is better, and ASR is opposite.}
\label{tab: different_trigger_pattern}
\resizebox{\linewidth}{!}{
\begin{tabular}{c|cccc}
\toprule
Attacks               & w/o   & \multicolumn{1}{c}{CA}                              & \multicolumn{1}{c}{PA}                               & \multicolumn{1}{c}{ASR}                            \\ \midrule
BadNets 5,5, grid     & 93.31 & 87.25   (\textcolor{red}{$\uparrow$0.12}) & 84.02 (\textcolor{dark_green}{$\downarrow$0.71})  & 0.60 (\textcolor{dark_green}{$\downarrow$1.78}) \\
BadNets, 3, 3, random & 92.98 & 86.31 (\textcolor{dark_green}{$\downarrow$0.82}) & 83.83 (\textcolor{dark_green}{$\downarrow$0.90})  & 1.20 (\textcolor{dark_green}{$\downarrow$1.18}) \\
TrojanNN\_FireFox     & 93.26 & 87.12 (\textcolor{dark_green}{$\downarrow$1.65}) & 83.81 (\textcolor{dark_green}{$\downarrow$0.91})  & 0.87 (\textcolor{dark_green}{$\downarrow$1.38}) \\
Blended\_Kitty2       & 93.28 & 87.30 (\textcolor{dark_green}{$\downarrow$1.22}) & 76.65 (\textcolor{red}{$\uparrow$1.96})    & 5.35 (\textcolor{red}{$\uparrow$1.34})   \\
SIG\_6\_20            & 93.38 & 88.86 (\textcolor{red}{$\uparrow$1.08})   & 77.41 (\textcolor{red}{$\uparrow$2.70})    & 2.76 (\textcolor{red}{$\uparrow$0.81})   \\
SIG\_3\_40            & 93.52 & 87.45 (\textcolor{dark_green}{$\downarrow$0.33}) & 59.12 (\textcolor{dark_green}{$\downarrow$15.59}) & 7.67 (\textcolor{red}{$\uparrow$5.72})  \\
\bottomrule
\end{tabular}
}
\end{table}

\section{Generalization for Different Trigger Patterns.}
In this section, we evaluate \sysname on different triggers to demonstrate the generalization, the corresponding results are presented in table~\ref{tab: different_trigger_pattern} where the colorful numbers and arrows represent the performance variations with respect to the CIFAR-10 in Sec.~\ref{subsec: purification_results}. Specifically, we replace the trigger pattern across three categories of trigger (high-visibility trigger and semi-visibility trigger). We select four representative attacks (BadNets, TrojanNN, Blended, and SIG) and replace their standard triggers with alternative patterns to generate new backdoored models. We do not perform this experiment for Low-Visibility Triggers (LVTs) because their triggers, such as the warping fields in WaNet or imperceptible perturbations in BPP, are not simple, replaceable patterns but rather complex, sample-specific transformations.

The results show that \sysname's performance is highly consistent across different trigger patterns, exhibiting only minor deviations from the baseline results. On average, the performance changes were minimal: a -0.50\% change in CA, a +1.44\% change in PA, and a -0.44\% change in ASR. The results indicate that the \sysname is generic for different types of triggers, rather than being limited to specific triggers. It is noteworthy that \sysname does not achieve particularly satisfactory results in the \textit{SIG\_3\_40} with its PA decreasing by 15.59\% compared to the baseline and its ASR increasing by 5.72\%. The reason is that the trigger increases/decreases pixel values by a maximum of \textit{40/255}, which means that, under ideal conditions, removing the backdoor would require a cost of approximately \textit{80/255}, severely damaging the original semantic information of the image. However, despite the significant decrease compared to the baseline, there is still a substantial improvement compared to SampDetox which PA is 4.01\% and ASR is 95.67\%. 

\begin{figure*}[!ht]
\centering
\includegraphics[width = 1\textwidth]{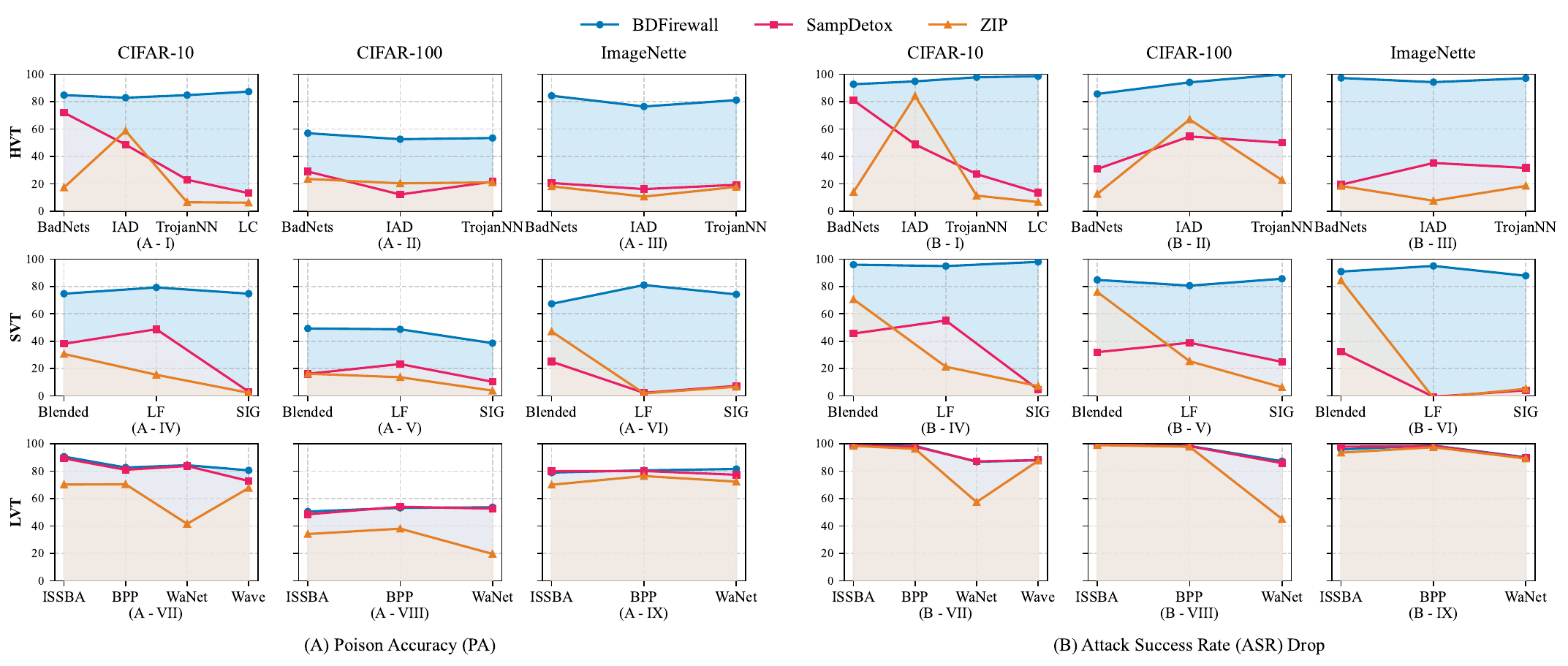}
\caption{Visualization of comparison results. We present them in two sub-figures, where sub-figure A reports the PA of three defense methods on different Attacks across three baseline datasets, and sub-figure B presents the ASR drop which is the decreasing value compare which the ASR of \textit{no defense}. For PA and ASR Drop, the higher the better, which means that methods that occupy a larger area in the graph have better defensive effects.}
\label{fig: comparison}
\end{figure*}

\section{Visualized performance comparison.}

Figure~\ref{fig: detail_comparison} visualizes the performance data from Tab.~\ref{tab: Performance comparison} to illustrate the core defense objectives: purifying backdoor samples to be inferred as their ground-truth labels on the backdoor model. We focus on two key metrics: PA and ASR Drop. Note that, the ASR drop, defined as the difference between the baseline ASR ('No Defense') and the post-defense ASR, is used to unify the comparison. Consequently, for both PA and ASR drop, higher values indicate superior performance. As the figure illustrates, our method (\sysname) consistently occupies a larger area on the chart, signifying its superior overall performance compared to the two other leading methods, especially against HVT- and SVT-based attacks. A detailed analysis of these results is presented in Sec.~\ref{subsec: purification_results}. To complement these results, we also provide a analysis of the purification process, offering visual evidence that explains the superior performance of \sysname.

\section{Details of Purification.}
\label{sec: Details of Purification}

\begin{figure*}[t]
    \centering
    \includegraphics[width=0.86\linewidth]{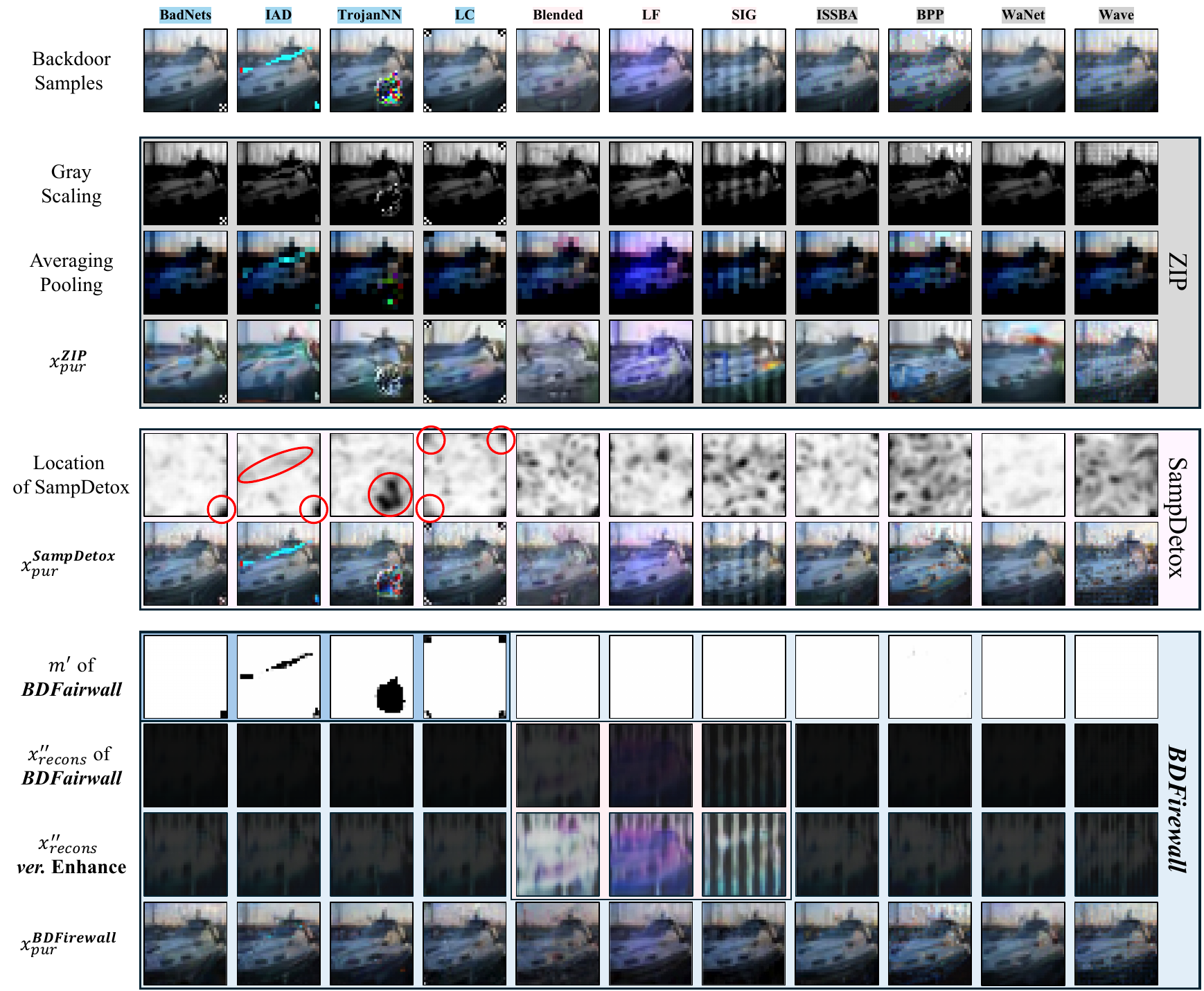}
    \caption{Visualization of Intermediate Purification Steps. Attacks are color-coded by trigger type: blue backgrounds (\colorbox{blue_in_fig1}{HVT}) for High-Visibility Trigger, pink (\colorbox{pink_in_fig1}{SVT}) for Semi-Visible Trigger, and gray (\colorbox{gray_in_fig1}{LVT}) for Low-Visibility Trigger. Each row visualizes the intermediate and final outputs for a specific defense method: ZIP, SampDetox, and \sysname.}
    \label{fig: detail_comparison}
    \vspace{-5 mm}
\end{figure*}

In this section, we visualize the purification processes of ZIP, SampDetox, and \sysname~to provide an intuitive analysis of their respective shortcomings and advantages. To provide a concrete example, we apply these methods to samples from the CIFAR-10 dataset. Specifically, ZIP's procedure includes two intermediate processes—gray-scaling and blurring (implemented via average pooling)—while SampDetox utilizes one intermediate process, which we term 'Location'. In contrast, the process of \sysname~also involves two key intermediate steps: locating the HVT and reconstructing the SVT. Accordingly, we illustrate the intermediate and final purification results of these methods in Figure~\ref{fig: detail_comparison}. 

\textbf{ZIP} \cite{ZIP_shi_2023} employs gray-scaling and blurring as linear transformations to destroy potential triggers, subsequently using a pre-trained diffusion model to restore the damaged image content. Ideally, the intermediate result of this process should be a trigger-free version of the original input. However, the visualizations in the second and third rows demonstrate that a simple global transformation not only fails to effectively destroy triggers—especially large patterns like TrojanNN and attacks based on SVT, but also inflicts irreversible damage on clean features. This leads to both incomplete trigger removal (e.g., \colorbox{blue_in_fig1}{BadNets}, \colorbox{blue_in_fig1}{TrojanNN}, \colorbox{pink_in_fig1}{LF}, \colorbox{pink_in_fig1}{SIG}) and collateral damage to the image's benign content (e.g., IAD, Blended, WaNet). The intrinsic reason for ZIP's failure lies in its reliance on a uniform, global purification strategy for all triggers, rather than adapting the destruction method to specific trigger characteristics. Consequently, this approach struggles to simultaneously achieve significant ASR reduction, preserve PA (PA), and maintain CA.

\textbf{SampDetox} \cite{SampDetox_yang_2024} leverages a 'Location' operation to identify HVT and purifies them by applying adaptive high-intensity noise to the identified regions. Ideally, this 'Location' operation should precisely identify triggers with minimal error. However, the visualizations reveal that the 'Location' results are far from accurate. For attacks based on SVT and LVT, SampDetox's design implies that they should be handled by a global purification mechanism. Consequently, the 'Location' operation is expected to find no specific trigger region, producing a nearly white output. However, the actual outputs are far from this expectation. Most of them neither located the trigger nor produced the expected white image, with the notable exception of \colorbox{gray_in_fig1}{WaNet}. As a consequence, balancing Prediction Accuracy (PA) and Attack Success Rate (ASR) for SVT and LVT attacks becomes challenging, a point further discussed in Sec.~\ref{subsec: purification_results}. The method fails to precisely isolate the trigger and, in turn, erroneously corrupts benign features. Furthermore, this inaccuracy extends to the noise addition of HVT purification, results in the HVT are hard to be purified (e.g., \colorbox{blue_in_fig1}{BadNets}, \colorbox{blue_in_fig1}{IAD}, \colorbox{blue_in_fig1}{TrojanNN}, and \colorbox{blue_in_fig1}{LC}). The root cause of these suboptimal outcomes is that SampDetox establishes a flawed correlation between model robustness and trigger visibility, which leads to misguided noise application.

Finally, \textbf{\sysname} leverages a progressive black-box framework to defend against the HVTs, SVTs and LVTs in successive order. In Stage I, we introduce a segmentation model to predict the location of HVTs, generating a binary mask denoted as $m^{\prime}$. Ideally, the black area of $m^{\prime}$ should precisely overlap with the HVT regions while remaining blank (i.e., white) for all other conditions. Stage II, in turn, focuses on reconstructing SVTs, producing an output denoted as $x^{\prime\prime}_{recons}$. The goal is to reconstruct the SVT pattern with minimal information loss while generating a black output for non-SVT inputs. Since Stage III is a straightforward noise-and-denoise process whose output is the final purified image, we do not visualize its intermediate results separately. 

The visualizations demonstrate that Stage I achieves excellent predictive accuracy, significantly outperforming SampDetox's 'Location' module. Specifically, the predicted mask $m^{\prime}$ accurately covers the trigger-patched areas for HVTs and remains correctly blank for SVT and LVT attacks. One notable exception occurs with the \colorbox{blue_in_fig1}{LC} attack: its black-and-white grid trigger was expected to produce a solid square mask in $m^{\prime}_{LC}$. However, the resulting mask $m^{\prime}_{LC}$ only captures the white portions of the trigger pattern. This discrepancy stems from our defense's core intuition: discriminating based on semantic differences between clean and poisoned features. Since the background pixels in that region are also black, there is no discernible semantic difference between the black parts of the trigger and the clean area. Consequently, our method only identifies the white trigger pixels that exhibit a clear semantic deviation. However, this partial detection does not impede trigger removal, as replacing black pixels with other black pixels has no effect on the outcome. This behavior, in fact, reflects that Stage I is operating precisely as designed.  In terms of Stage II, we present both the original reconstructed output ($x^{\prime\prime}_{recons}$) and an enhanced version ($x^{\prime\prime}_{recons}$ \textit{ver. Enhance}) created by magnifying the original output threefold for better visibility. Stage II successfully reconstructs various SVTs—such as the Hello Kitty pattern in \colorbox{pink_in_fig1}{Blended}, the purple region in \colorbox{pink_in_fig1}{LF}, and the sinusoidal signal in \colorbox{pink_in_fig1}{SIG}—while keeping the output for other attacks predominantly black. Note that, although the reconstruction model was trained with constraints, it cannot produce a perfect zero output (i.e., pure black), thus naturally introducing a slight bias. This explains why Stage II can cause minor degradation to clean features. However, compared to the baseline methods, our approach not only removes SVTs more effectively but also significantly reduces collateral damage to benign features.

In terms of Stage II, we present the origin results ($x^{\prime\prime}_{recons}$) and we also present the corresponding enhanced version ($x^{\prime\prime}_{recons}$ \textit{ver. Enhance}) by directly magnifying $x^{\prime\prime}_{recons}$ three times. We can see that Stage II effectively reconstruct the SVTs (the hellokitty of \colorbox{pink_in_fig1}{Blended}, the purple area of \colorbox{pink_in_fig1}{LF} and the sinusoidal signal of \colorbox{pink_in_fig1}{SIG}) and kept black as much as possible for other attacks. Note that, although we restricted $f^{\prime\prime}$ during training, since a model cannot output 0 (i.e., pure black) completely, it naturally introduces bias. This explains why the introduction of Stage II introduces some damage to clean features. However, compared with the two baseline methods, our method not only effectively removes SVTs but also significantly reduces damage to clean features.

In summary, we have presented a detailed analysis of the intermediate outputs generated during the purification processes of ZIP, SampDetox, and \sysname. These visualizations elucidate why prior methods falter, while simultaneously demonstrating how \sysname can effectively and efficiently neutralize such backdoors. Furthermore, these visual results provide compelling, intuitive evidence that corroborates our quantitative experimental findings, confirming both the accuracy of our analysis and the superiority of our proposed method.

\end{document}